\documentclass[aps,prm,twocolumn,superscriptaddress]{revtex4-1}
\usepackage{graphicx,bm,amssymb,amsmath,xcolor,pifont,soul}
\usepackage[linktocpage=true,colorlinks=true,pdfborder={0 0 0},linkcolor=blue,citecolor=red,filecolor=yellow,urlcolor=blue,bookmarks,pdfauthor={},]{hyperref}
\usepackage{ulem}
\usepackage{multirow}
\usepackage[T1]{fontenc}

\begin{document}

%\title{Suppression of charge-density wave and enhancement of superconductivity in Li-intercalated NbSe$_2$}
\title{Suppression of charge-density wave and superconductivity in a
lithiated NbSe$_2$ monolayer}
\author{Hari Paudyal}
\email{ph.paudyal@gmail.com}
\affiliation{Department of Physics and Astronomy, University of Iowa, Iowa City, Iowa 52242, USA}
\author{Michael E. Flatt\'e}
\email{michaelflatte@quantumsci.net}
\affiliation{Department of Physics and Astronomy, University of Iowa, Iowa City, Iowa 52242, USA}
\affiliation{Department of Applied Physics and Science Education, Eindhoven University of Technology, Eindhoven 5612 AZ, The Netherlands}

\date{\today}

\begin{abstract}
We present an \textit{ab initio} investigation of the long-range charge density wave (CDW) order and superconducting properties of the pristine and lithiated NbSe$_2$ monolayer. Stable CDW structures are obtained through atomic reconstruction driven by soft-mode distortions and lithiation, respectively, lead to significant electronic modifications that suppress the CDW order. This suppression is attributed to anisotropic atomic distortions, along with a reduction in the electronic density of states at the Fermi level. As a result, the electron--phonon coupling strength is suppressed, particularly in the lithiated structure, due to reduced contributions from low-frequency phonons, primarily associated with in-plane Nb vibrations. Finally, we observe a sizable anisotropy in the superconducting gap on the Fermi surface, with a superconducting transition temperature of approximately 8~K in the distorted, and 4~K in the lithiated, CDW NbSe$_2$ monolayer.

\end{abstract}
\maketitle
%\section{Introduction}
Transition metal dichalcogenides (TMDs) have been proven to be a highly versatile class of van der Waals (vdW) layered materials, offering rich electronic properties suited for applications in nanoelectronics and spintronics~\cite{xu2014, chhowalla2016, mak2016, manzeli2017}. Notable examples include MoS$_2$, which has been extensively studied for its semiconducting properties and potential applications in field-effect transistors, and WS$_2$, known for its strong spin-orbit coupling and pronounced photoluminescence in the monolayer form~\cite{zhu2011, zhao2013}. Similarly, MoTe$_2$ and WTe$_2$ display large nonsaturating magnetoresistance and pressure-driven superconductivity, and have been predicted to be type-II Weyl semimetals~\cite{ali2014, qi2016, deng2016, wang2016, paudyal2020}, while TiSe$_2$, NbS$_2$, and NbSe$_2$ have been actively explored for their charge density wave (CDW) phenomena and  superconductivity~\cite{morosan2006, rossnagel2011, soumyanarayanan2013, xi2015cdw}. Since NbSe$_2$ is a prototypical candidate exhibiting both CDW order (below 33~K) and superconductivity (below 7.2~K)~\cite{xi2015cdw, soumyanarayanan2013, xi2016ising, yokoya2001}, it is essential to understand the interplay between the superconducting pairing mechanism, the Fermi surface (FS) topology, and the structural modulation associated with the CDW state~\cite{kiss2007, borisenko2009}. In addition to conventional superconductivity, monolayer NbSe$_2$ has also been identified as a model system for Ising-type superconductivity, where strong spin-orbit coupling locks electron spins out of plane, protecting Cooper pairs against in-plane magnetic fields~\cite{xi2016ising, Das2022s}.

Recent advances in angle-resolved photoemission spectroscopy, scanning tunneling microscopy, and first-principles calculations have provided deeper insights into the electronic structure of the NbSe$_2$ monolayer, revealing how subtle changes in the FS and lattice distortions contribute to the emergence of a CDW and superconductivity~\cite{xi2015cdw, nakata2018anisotropic, kundu2024cdw}. The CDW phase is characterized by a pronounced soft phonon mode around the wave vector \textbf{q}$_{\rm CDW} = 2/3\Gamma M$, indicating a strong structural instability that drives the formation of a $3 \times 3$ real space supercell, resulting from Nb-Nb clustering~\cite{Weber2011s, xi2015cdw}. This soft phonon mode has been attributed to strong e-ph coupling, which plays a key role not only in the superconducting pairing mechanism but also in stabilizing the CDW phase~\cite{kiss2007, zheng2019s}. However, the microscopic origin of this coupling remains unresolved, and its relationship with superconductivity continues to be  a subject of debate~\cite{borisenko2009, kundu2024cdw}. On the other hand, a few strategies have manipulated the CDW order, including  applying pressure~\cite{suderow2005}, chemical doping, and intercalation~\cite{calandra2011}. These approaches show that the interplay between CDW and superconductivity does not follow a universal trend with a straightforward relationship. For instance, whereas external pressure suppresses the CDW phase and enhances superconductivity, the intercalation of alkali metals (like lithium) or application of strain~\cite{kundu2024cdw}  destabilizes the CDW phase, leading to an increased superconducting transition temperature~\cite{morosan1991, calandra2011}. Such tunability of the CDW phase not only deepens our understanding of the competition between these two low-temperature phases but also opens pathways for designing materials with tailored properties for low-temperature electronics, as well as other technological applications~\cite{nakata2018anisotropic}.

In this study, we perform \textit{ab initio} calculations to investigate the atomic structure and lattice dynamics of the distorted and lithiated CDW phases, as well as the influence of CDW-induced electronic modulations on the superconducting properties of the NbSe$_2$ monolayer. All calculations are carried out within the density functional theory framework, employing the generalized gradient approximation for the exchange-correlation functional. Structural relaxation, electronic structure, and phonon calculations are performed using the \textsc{Quantum ESPRESSO} package~\cite{QE_methodology}. The electron-phonon (e-ph) matrices and superconducting properties are subsequently evaluated using the \textsc{EPW} code~\cite{EPW_methodology}, based on density functional perturbation theory and Wannier interpolation techniques. Maximally localized Wannier functions are constructed using the \textsc{Wannier90} code~\cite{W90_methodology}, enabling efficient interpolation of electronic bands and e-ph matrix elements on extra dense \textbf{k}/\textbf{q} grids.

Our results confirm the formation of a stable $3 \times 3$ supercell in the CDW phase, driven primarily by Nb-Nb clustering and strong phonon softening near the $M$ point. The distorted structure exhibits a lower total energy compared to the undistorted phase, with the elimination of imaginary phonon modes, confirming its dynamical stability. Upon lithium adsorption we observe similar structural distortions, however, the electronic structure differs significantly from that of the pristine CDW phase. As lithium acts as an electron donor it induces an upward shift of the Fermi level ($E_{\rm F}$), modulating the charge distribution and suppressing the CDW transition. However, this suppression is accompanied by a reduction in the e-ph coupling strength associated primarily from the low-energy phonons, leading to a decrease in the superconducting transition temperature ($T_c$). These results clarify how lithiation affects the balance between CDW order and superconductivity in NbSe$_2$ monolayer, highlighting its potential as a practical route to tune phases in low-dimensional materials.

\begin{figure}[t!]
\centering
\includegraphics[width=0.43\textwidth]{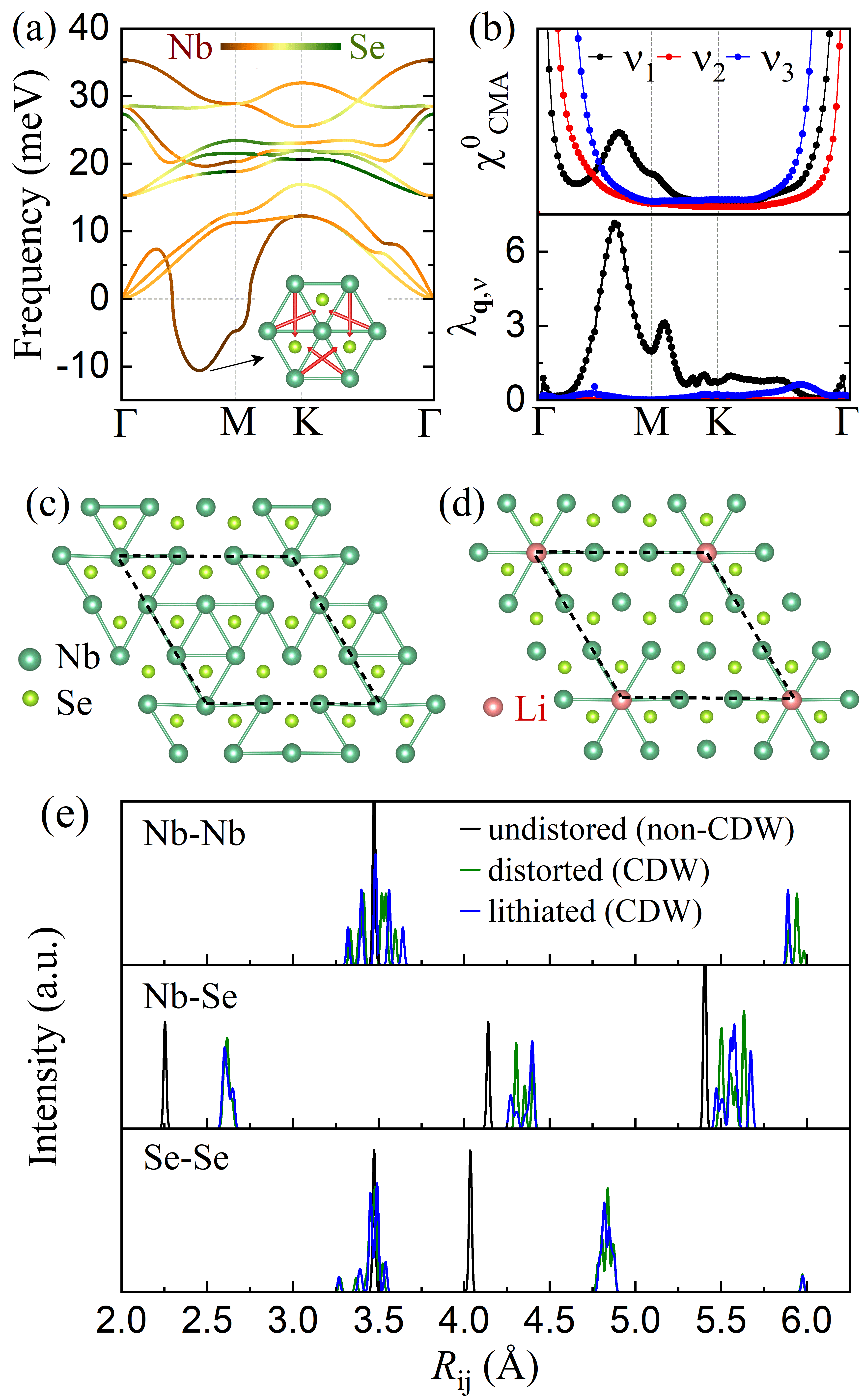}\hfill
\caption{(a) Phonon dispersion of monolayer NbSe$2$ showing an imaginary mode along $\Gamma-M$ direction, indicating a CDW instability. Color scale represents atomic contributions to the phonon modes. Inset: top view of the pristine structure with Nb displacements corresponding to the soft mode. (b) Static electronic susceptibility $\chi^0{\rm CMA}$ (top) and mode-resolved e-ph coupling strength $\lambda_{\mathbf{q}\nu}$ (bottom), both peaking at the \textbf{q}$_{\rm CDW}$. Fully relaxed structure of the (c) distorted CDW and (d) lithiated CDW phases, highlighting Nb-Nb bonds of 3.5-3.6~\AA. (e) Partial radial distribution functions $R_{ij}$ for Nb-Nb, Nb-Se, and Se-Se in the undistorted (black), distorted CDW (red), and lithiated CDW (blue) structures. The CDW formation in both cases lead to distinct bond-length modulations and local distortions in the system.}
\label{fig1}
\end{figure}

\begin{figure*}[t]
\centering
\includegraphics[width=0.90\textwidth]{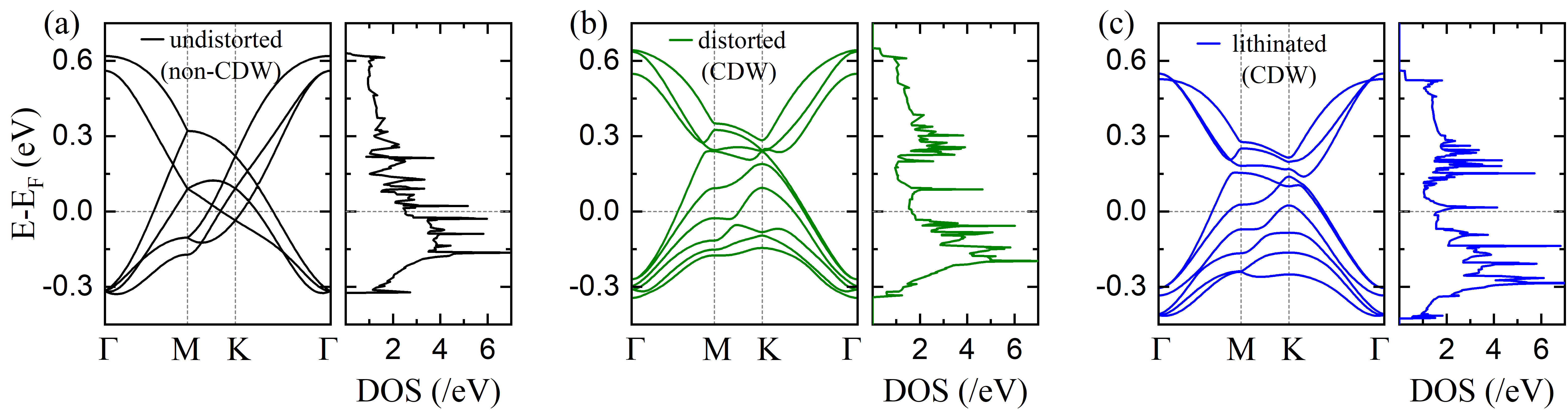}\hfill
\caption{Electronic band structure and DOS (states/eV/u.c.) for (a) undistorted, (b) distorted CDW, and (c) lithiated CDW structures.}
\label{fig2}
\end{figure*}

NbSe$_2$ crystallizes in a hexagonal structure with a double-layered unit cell (space group $P6_3/mmc$), where the layers are weakly bonded through vdW interactions~\cite{wilson1975}. Atomically flat and chemically stable monolayers can be obtained via mechanical exfoliation~\cite{novoselov2005mechanical}, consisting of a plane of Nb atoms sandwiched between two planes of Se atoms. Each Nb atom is coordinated by six nearest-neighbor Se atoms, forming a trigonal prismatic geometry with the Nb-Se bond length of $2.615$~\AA, characteristic of the hexagonal close-packed structure. The relaxed in-plane lattice constant is found to be $3.473$~\AA~corresponding to the Nb-Nb bond length, in good agreement with previous experimental and theoretical studies~\cite{nakata2018anisotropic}. The phonon dispersion of the NbSe$_2$ monolayer calculated within the harmonic approximation (Fig.~\ref{fig1}(a)), shows that one of the lowest-energy acoustic vibrational modes exhibits imaginary frequencies along the 2/3$\Gamma M$ direction, indicating a structural instability~\cite{Weber2011s}. This mode is mainly governed by in-plane vibrations of the Nb atoms which are not strictly alternating but are localized and directionally dependent (inset of Fig.\ref{fig1}(a)) and reflects the onset of a commensurate $3 \times 3$ CDW real-space superstructure, as confirmed by scanning tunneling microscopy measurements~\cite{Yoshizawa2024, Sivakumar2025}, and  first principles calculations~\cite{zheng2018}.

In particular, the softening of low-energy phonon modes and the lifting of electronic degeneracies due to phonon-induced lattice instabilities are hallmark signatures of a momentum-dependent e-ph coupling-driven CDW instability~\cite{johannes2008fermi, calandra2011cdw}. To investigate the mechanism underlying the CDW transition in monolayer NbSe$_2$, we first compute the static electronic susceptibility, $\chi^0{\text{CMA}}$ --- the purely electronic contribution --- using the constant matrix approximation, followed by the calculation of the e-ph coupling strength, $\lambda_{\mathbf{q},\nu}$, for the three low-energy modes~\cite{ying2018unusual}. As shown in Fig.\ref{fig1}(b), $\chi^0_{\text{CMA}}$ exhibits a modest peak, whereas $\lambda_{\mathbf{q},\nu}$ displays pronounced maxima corresponding to the longitudinal acoustic mode with imaginary frequency at the $\mathbf{q}_{\text{CDW}}$. These results indicate that the electronic susceptibility alone is insufficient to induce the CDW transition; rather, it acts synergistically with strong e-ph coupling at the $\mathbf{q}_{\text{CDW}}$ to stabilize the CDW phase~\cite{johannes2008fermi}.

The stable CDW phases are constructed within a $3 \times 3$ supercell using two distinct structural models: (i) the distorted CDW structure, generated by displacing atoms along the eigenvectors of the imaginary phonon mode at the $\mathbf{q}_{\text{CDW}}$, and (ii) the lithiated CDW structure, obtained by placing a Li atom directly above a Nb site. In the first approach, we initiate the CDW transition and hence symmetry breaking by following a physically motivated distortion pathway derived from lattice dynamics, in contrast to earlier studies that employed random atomic displacements~\cite{lian2018, zheng2018}. These structures are subsequently relaxed using high-convergence criteria to ensure residual atomic forces are minimized below 1~meV/\AA. 

The relaxed distorted CDW structure is energetically favored over the undistorted phase by $\sim$50~meV per atom, whereas the lithiated CDW structure --- with Li placed above the Nb site --- is more stable than the configuration with Li at the hollow site by $\sim$44~meV/atom~\cite{formation_energy}. In both structures, the lattice distortion modulates Nb-Nb bond lengths, forming distinct Nb clusters (Fig.\ref{fig1}(c) and (d)) --- another structural hallmark of the CDW modulation, consistent with previous reports~\cite{lian2018, zheng2018}. The Nb-Nb distances range from 3.33 to 3.60~\AA, reflecting symmetry breaking. whereas the Nb-Se and Se-Se bond lengths vary from 2.59 to 2.66~\AA~and 3.44 to 3.53~\AA, respectively, indicating subtle out-of-plane and lateral relaxations induced by the CDW distortion. %It is to be noted that the formation of the CDW structure is evident not only through the reconstruction of the Nb atomic layer, which occurs alongside the small but not negligible displacement of the Se atoms.

These structural modulations induced by the CDW transition are further confirmed by calculated radial distribution functions (RDFs) (Fig.~\ref{fig1}(e)). A pronounced deviation in the structural modulation is observed in the first-nearest-neighbor peaks, particularly for Nb-Nb and Nb-Se pairs in both CDW structures. To further quantify structural similarities, we compute a similarity factor (SF) defined as the dot product between RDFs. The SF between the undistorted and distorted CDW structures is 11\%, and between the undistorted and lithiated CDW structures is 8\%, indicating that both CDW configurations are structurally distinct from the undistorted phase. Interestingly, the SF between the distorted and lithiated CDW phases is 67\%, suggesting that although they share common features --- particularly in the first-nearest-neighbor distances --- they indeed remain structurally distinct. The primary differences arise from the second-nearest-neighbor shell, which is expected to influence both the electronic and vibrational properties. These distortions are characteristic of CDW ordering and lead to a local symmetry breaking, resulting in notable changes in the electronic DOS at $E_{\rm F}$ and the splitting of phonon modes.% due to reduced lattice symmetry.

Figure~\ref{fig2} compares the electronic band structure and density of states (DOS) for monolayer NbSe$_2$ in the undistorted, distorted CDW, and lithiated CDW structures. The most prominent signature of the CDW transition is the reconstruction of the band structure arising from periodic lattice distortions, which leads to the splitting of bands near $E_{\rm F}$~\cite{johannes2008fermi}. Notably, whereas the orbital character of the bands remains largely unchanged, the bandwidths broaden, as evidenced by the steeper slopes of the dispersion splitting associated with CDW-induced symmetry breaking. This splitting reflects a redistribution of electronic states, lowering the energy of occupied bands and contributing to the energetic stabilization of the CDW phase~\cite{Weber2011s}. These effects are particularly evident near the M and K points of the Brillouin zone, where the bands exhibit flattening and partial gap openings of approximately 100-200~meV. As expected, the enhanced CDW distortion results in a pronounced reduction in the DOS at $E_{\rm F}$. A similar electronic reconstruction in the distorted CDW phase is also reported in the work of Lian~\textit{et al.}~\cite{lian2018}, showing excellent agreement with the present calculations. Lithium adsorption introduces additional charge carriers, shifting $E_{\rm F}$ upward and reducing the DOS at $E_{\rm F}$, suggesting suppression of the CDW order due to electron doping~\cite{calandra2011cdw, zheng2019s}. Additionally, the band structure reconstruction leads to notable modifications in FS, consistent with the reduced nesting observed in the distorted CDW phase~\cite{johannes2008fermi}.

\begin{figure}[!]
\centering
\includegraphics[width=0.42\textwidth]{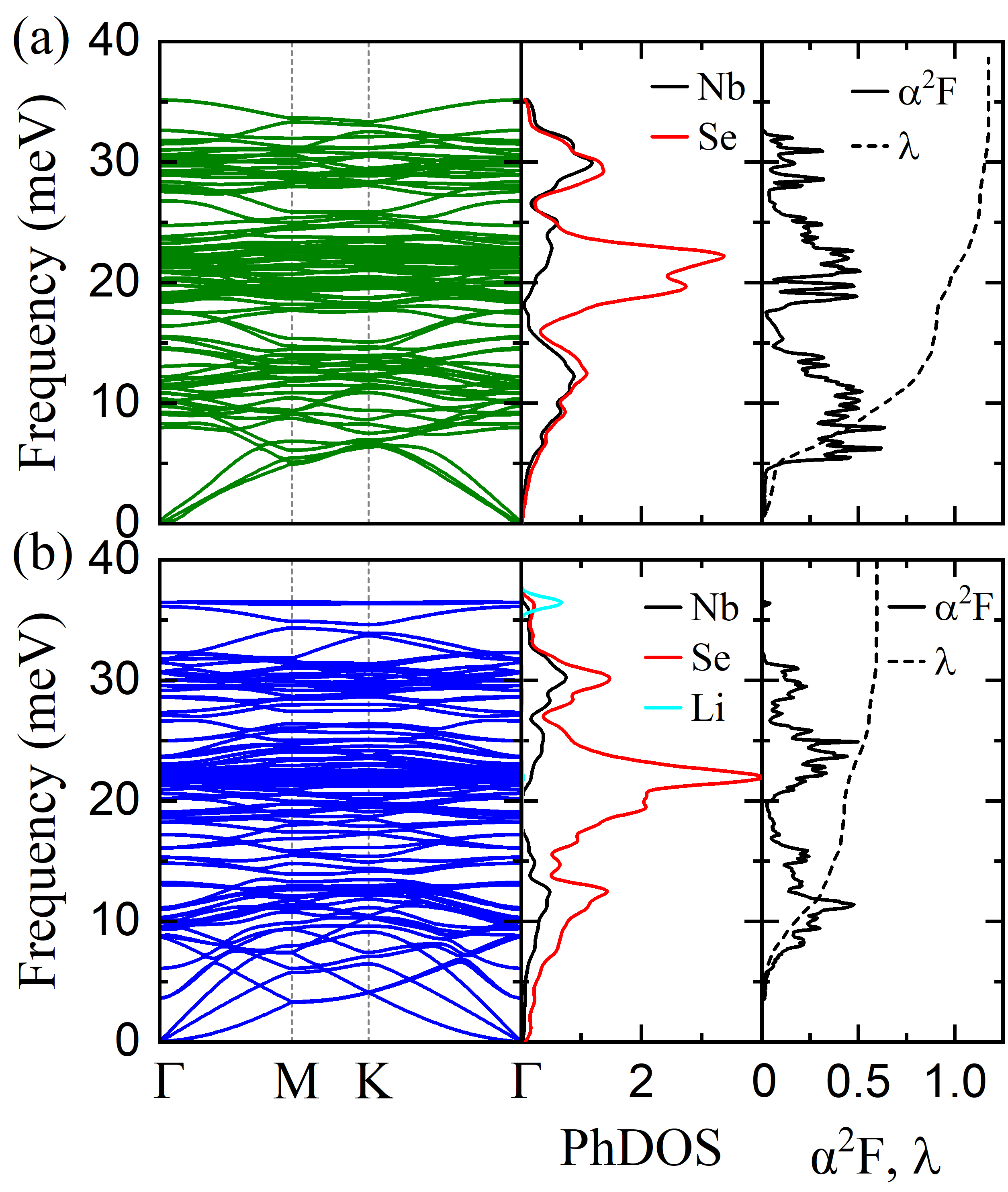}\hfill
\caption{Phonon dispersion, atom projected phonon DOS, Eliashberg spectral function $\alpha^2F$, and cumulative e-ph coupling strength $\lambda$ for (b) distorted (CDW) and (c) lithiated CDW structures.}
\label{fig3}
\end{figure}

Figure~\ref{fig3} shows the calculated phonon dispersion, projected phonon density of states (PhDOS), Eliashberg spectral function $\alpha^2F(\omega)$, and cumulative e-ph coupling strength $\lambda(\omega)$ for the (a) distorted and (b) lithiated CDW phases. The absence of imaginary frequencies in both structures confirms their dynamic stability. Compared to the undistorted structure, the phonon frequencies in the CDW phase are generally higher, and phonon degeneracies are lifted due to symmetry breaking induced by lattice distortion. This symmetry reduction leads to the splitting of phonon branches, especially in the low-frequency acoustic region. In the distorted CDW phase, the atom projected PhDOS reveals that the low-frequency modes below 15~meV are predominantly composed of in-plane vibrations of Nb atoms. These modes exhibit strong coupling to the electronic states, as reflected in the enhanced values of $\alpha^2F(\omega)$ in the same energy range. In contrast the mid- and high-frequency regions (above 20~meV) contain vibrational contributions primarily from Se atoms and out-of-plane Nb motions, which contribute relatively less to the total e-ph coupling. In the lithiated CDW structure, the intensity of low-frequency Nb vibrations is notably reduced, consistent with the observed weakening of e-ph coupling in this energy range. Further, our harmonic phonon calculations reveal discrepancies in the low-energy acoustic modes below 6~meV when compared to previous studies~\cite{lian2018, zheng2019s}. These differences may arise from the fine structural variations and used computational settings, such as force constant convergence, choice of the pseudopotentials, or structural relaxation criteria-highlighting the sensitivity of phonon spectrum to subtle methodological choices. Further analysis of Raman-active modes in the distorted CDW phase identifies two high-frequency CDW-induced modes with $A_1$ symmetry at 81.2~cm$^{-1}$ and 192.1~cm$^{-1}$, in reasonable agreement with experimental observations and prior theoretical results~\cite{xi2015cdw, lian2018}. These modes are dominated by in-plane vibrations of Nb atoms, as previously reported~\cite{lian2018}, but also exhibit notable contributions from the Se vibrations, highlighting their role in CDW stabilization. Further, in the lithiated CDW phase, these modes are found at 81.7~cm$^{-1}$ and 191.1~cm$^{-1}$, indicating a slight redshift in the higher-frequency mode due to Li-induced lattice distortion.

\begin{figure}[t!]
\centering
\includegraphics[width=0.42\textwidth]{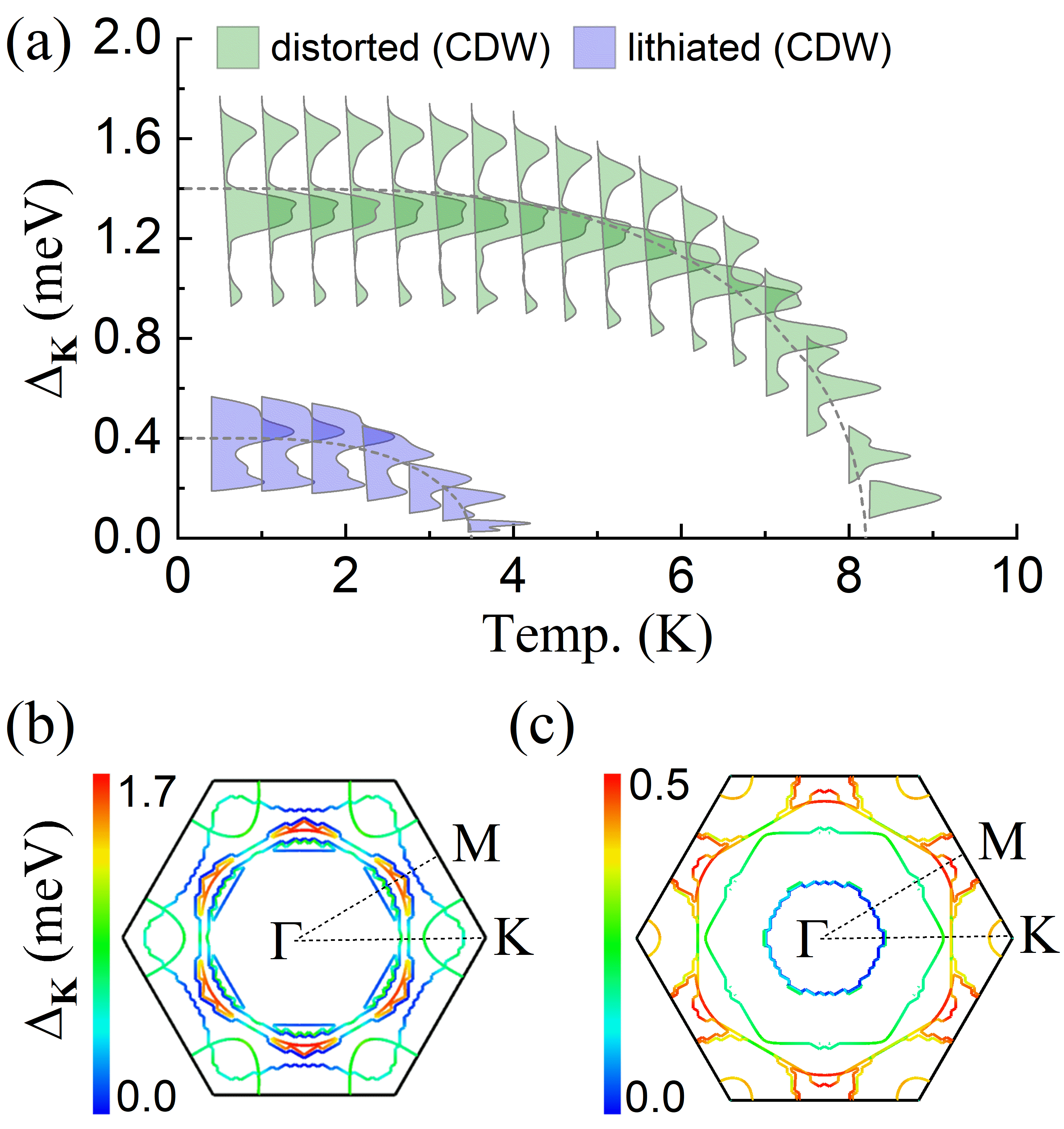}\hfill
\caption{(a) Temperature dependence of the superconducting gap distribution $\Delta_{\mathbf{k}}$ for the distorted (green) and lithiated (blue) CDW phases of monolayer NbSe$2$. The shaded regions represent the anisotropic spread of $\Delta{\mathbf{k}}$ across FS, and the dashed lines are guides to the eye obtained by fitting the BCS gap equation using the calculated corresponding value of $\lambda$. FS maps of the superconducting gap $\Delta_{\mathbf{k}}$ at low temperature (1~K) for the (b) distorted and (c) lithiated CDW phases. A sizable gap anisotropy and multigap features are evident in the distorted phase, whereas a reduced and more isotropic gap structure is observed in the lithiated phase.}
\label{fig4}
\end{figure}

We further examine the isotropic Eliashberg spectral function $\alpha^2F(\omega)$ and cumulative e-ph coupling constant $\lambda(\omega)$, in order to investigate the nature of the e-ph coupling and hence the superconducting properties of the stable CDW phases (right panels Fig.~\ref{fig3}). A comparison of the $\alpha^2F(\omega)$ with the PhDOS reveals enhanced coupling to low-frequency vibrational modes below 15~meV in the distorted CDW, which contributes approximately 65\% of the total $\lambda = 1.2$. In contrast, in the lithiated CDW, these modes exhibit weaker coupling, accounting for only about 40\% of the total $\lambda = 0.7$. The values of $\lambda$ reported in previous studies for the distorted CDW structure are notably underestimated and show inconsistencies, particularly in their treatment of the low-energy phonon contributions despite agreement in structural and electronic properties.~\cite{lian2018, zheng2019s}. The strong coupling observed in our distorted CDW phase highlights the sensitivity of superconducting properties to the phonon dispersion and accurate treatment of phonons particularly in the low-energy region.
%critical role of low-frequency phonons in mediating e-ph interactions.
%Interestingly, both studies report significantly lower $\lambda$ values compared to the present work, despite agreement in structural and electronic properties. This discrepancy highlights the sensitivity of superconducting properties to the phonon dispersion and accurate treatment of low-energy modes.

We now focus on the superconducting properties of the distorted CDW phases. The anisotropic superconducting gap $\Delta_{\mathbf{k}}$ is calculated by solving the gap equations self-consistently~\cite{Margine2013s}. Figure~\ref{fig4}(a) presents the anisotropic energy distribution of $\Delta_{\text{n}\mathbf{k}}$ as a function of temperature, evaluated with an effective Coulomb repulsion $\mu^* = 0.2$. A continuous, anisotropic superconducting gap emerges on the FS (Fig.~\ref{fig4}(b) and (c)), providing clear evidence of a multi-gap structure. Notably, in the distorted CDW phase, a prominent peak appears in the energy distribution around 1.2-1.3~meV, accompanied by two smaller peaks near 1.0~meV and 1.6~meV, indicating significant anisotropy in $\Delta_{\text{n}\mathbf{k}}$. Since the $\lambda$ is significantly reduced in the lithiated CDW phase, the $\Delta_{\text{n}\mathbf{k}}$ is also suppressed, with its energy distribution reduced to less than half of that in the distorted CDW phase. As the temperature increases, the superconducting gap gradually decreases and eventually vanishes at the critical temperature $T_{\text{c}}$, estimated to be 8.2~K for the distorted CDW structure and 3.8~K for the lithiated CDW structure. Anisotropy in the superconducting gap for the distorted CDW structure has also been reported in a previous study by Zheng~\textit{et al.}~\cite{zheng2019s}, although their calculated energy distribution is underestimated due to weak coupling contributed by low-frequency phonons, however, the multi-gap nature is qualitatively retained.%, and they estimated a critical temperature $T_{\text{c}}$ of approximately 4.4~K. Similarly, Lian~\textit{et al.}~\cite{lian2018} reported a lower $T_{\text{c}}$ of around 4.5~K by solving the Allen-Dynes-modified McMillan equation.% with a $\lambda = 0.84$.

An important open question in the community is why first-principles calculations consistently overestimate the critical temperature and zero-temperature superconducting gap especially in the TMDs~\cite{paudyal2020}. In the CDW phase, we find that this discrepancy may stem from the low-lying phonons which exhibit strong coupling. Further, there are a few limitations of the Migdal-Eliashberg formalism in capturing strong electronic correlations or anharmonic phonon effects, for instance, the choice of $\mu^*$, which introduces uncertainty in the estimated $T_{\text{c}}$. In such systems, very strong repulsive interactions at a few FS hot spots result in a relatively high value of $\mu^*$~\cite{Heil2017s}. Despite employing a relatively high value of $\mu^* = 0.2$, our calculated $T_{\text{c}}$ of 8.2~K for the distorted CDW structure remains slightly higher than experimental values, whereas previous studies employ $\mu^* = 0.1$ and estimate $T_{\text{c}}$ of around 4~K~\cite{lian2018, zheng2019s}. Overall, these findings suggest additional effects --- such as many-body renormalization or phonon softening beyond the harmonic approximation --- may play an important role in describing superconductivity in the NbSe$_2$ monolayer. 

In this work, we systematically investigate the structural, vibrational, and superconducting properties of the distorted and lithiated CDW phases of NbSe$_2$ monolayer using first-principles calculations. The stable $3 \times 3$ distorted CDW structure is achieved by displacing atoms along the eigenvectors of the soft phonon mode, and the lithiated structure is obtained by Li adsorption, with both structures subsequently relaxed using high-precision structural optimization. This distortion leads to distinct Nb clustering and substantial changes in the electronic structure, characterized by the removal of band degeneracies and a reduction in the density of states at $E_{\rm F}$. Phonon dispersion and Eliashberg spectral function calculations reveal strong e-ph coupling in the low-frequency region, with a total $\lambda = 1.2$ in the distorted CDW phase, contributing to a multigap superconducting state with an $T_{\text{c}}$ of 8.2~K. In contrast, Li adsorption weakens the CDW order, reduces $\lambda$ to 0.7, and suppresses superconductivity, yielding a lower $T_{\text{c}}$ of 3.8~K. Our findings demonstrate that both structural distortions and low-energy phonons play a central role in the e-ph coupling and hence  superconducting behavior of NbSe$_2$ monolayer, and lithiation provides an effective means to tune the balance between CDW and superconducting phases, offering promising implications for phase engineering in low-dimensional materials.

This work is supported by the U.S. Department of Energy, Office of Science, Office of Basic Energy Sciences under Award Number DE-SC0016379. We acknowledge use of the computational facilities at the University of Iowa (Argon cluster) and the Texas Advanced Computing Center (Frontera: DMR23051).

\bibliography{references}

\begin{thebibliography}{51}%
\makeatletter
\providecommand \@ifxundefined [1]{%
 \@ifx{#1\undefined}
}%
\providecommand \@ifnum [1]{%
 \ifnum #1\expandafter \@firstoftwo
 \else \expandafter \@secondoftwo
 \fi
}%
\providecommand \@ifx [1]{%
 \ifx #1\expandafter \@firstoftwo
 \else \expandafter \@secondoftwo
 \fi
}%
\providecommand \natexlab [1]{#1}%
\providecommand \enquote  [1]{``#1''}%
\providecommand \bibnamefont  [1]{#1}%
\providecommand \bibfnamefont [1]{#1}%
\providecommand \citenamefont [1]{#1}%
\providecommand \href@noop [0]{\@secondoftwo}%
\providecommand \href [0]{\begingroup \@sanitize@url \@href}%
\providecommand \@href[1]{\@@startlink{#1}\@@href}%
\providecommand \@@href[1]{\endgroup#1\@@endlink}%
\providecommand \@sanitize@url [0]{\catcode `\\12\catcode `\$12\catcode `\&12\catcode `\#12\catcode `\^12\catcode `\_12\catcode `\%12\relax}%
\providecommand \@@startlink[1]{}%
\providecommand \@@endlink[0]{}%
\providecommand \url  [0]{\begingroup\@sanitize@url \@url }%
\providecommand \@url [1]{\endgroup\@href {#1}{\urlprefix }}%
\providecommand \urlprefix  [0]{URL }%
\providecommand \Eprint [0]{\href }%
\providecommand \doibase [0]{http://dx.doi.org/}%
\providecommand \selectlanguage [0]{\@gobble}%
\providecommand \bibinfo  [0]{\@secondoftwo}%
\providecommand \bibfield  [0]{\@secondoftwo}%
\providecommand \translation [1]{[#1]}%
\providecommand \BibitemOpen [0]{}%
\providecommand \bibitemStop [0]{}%
\providecommand \bibitemNoStop [0]{.\EOS\space}%
\providecommand \EOS [0]{\spacefactor3000\relax}%
\providecommand \BibitemShut  [1]{\csname bibitem#1\endcsname}%
\let\auto@bib@innerbib\@empty
%</preamble>
\bibitem [{\citenamefont {Xu}\ \emph {et~al.}(2014)\citenamefont {Xu}, \citenamefont {Yao}, \citenamefont {Xiao},\ and\ \citenamefont {Heinz}}]{xu2014}%
  \BibitemOpen
  \bibfield  {author} {\bibinfo {author} {\bibfnamefont {X.}~\bibnamefont {Xu}}, \bibinfo {author} {\bibfnamefont {W.}~\bibnamefont {Yao}}, \bibinfo {author} {\bibfnamefont {D.}~\bibnamefont {Xiao}}, \ and\ \bibinfo {author} {\bibfnamefont {T.~F.}\ \bibnamefont {Heinz}},\ }\href {\doibase 10.1038/nphys2942} {\bibfield  {journal} {\bibinfo  {journal} {Nature Physics}\ }\textbf {\bibinfo {volume} {10}},\ \bibinfo {pages} {343} (\bibinfo {year} {2014})}\BibitemShut {NoStop}%
\bibitem [{\citenamefont {Chhowalla}\ \emph {et~al.}(2016)\citenamefont {Chhowalla}, \citenamefont {Jena},\ and\ \citenamefont {Zhang}}]{chhowalla2016}%
  \BibitemOpen
  \bibfield  {author} {\bibinfo {author} {\bibfnamefont {M.}~\bibnamefont {Chhowalla}}, \bibinfo {author} {\bibfnamefont {D.}~\bibnamefont {Jena}}, \ and\ \bibinfo {author} {\bibfnamefont {H.}~\bibnamefont {Zhang}},\ }\href {\doibase 10.1038/natrevmats.2016.52} {\bibfield  {journal} {\bibinfo  {journal} {Nature Reviews Materials}\ }\textbf {\bibinfo {volume} {1}},\ \bibinfo {pages} {16052} (\bibinfo {year} {2016})}\BibitemShut {NoStop}%
\bibitem [{\citenamefont {Mak}\ and\ \citenamefont {Shan}(2016)}]{mak2016}%
  \BibitemOpen
  \bibfield  {author} {\bibinfo {author} {\bibfnamefont {K.~F.}\ \bibnamefont {Mak}}\ and\ \bibinfo {author} {\bibfnamefont {J.}~\bibnamefont {Shan}},\ }\href {\doibase 10.1038/nphoton.2015.282} {\bibfield  {journal} {\bibinfo  {journal} {Nature Photonics}\ }\textbf {\bibinfo {volume} {10}},\ \bibinfo {pages} {216} (\bibinfo {year} {2016})}\BibitemShut {NoStop}%
\bibitem [{\citenamefont {Manzeli}\ \emph {et~al.}(2017)\citenamefont {Manzeli}, \citenamefont {Ovchinnikov}, \citenamefont {Pasquier}, \citenamefont {Yazyev},\ and\ \citenamefont {Kis}}]{manzeli2017}%
  \BibitemOpen
  \bibfield  {author} {\bibinfo {author} {\bibfnamefont {S.}~\bibnamefont {Manzeli}}, \bibinfo {author} {\bibfnamefont {D.}~\bibnamefont {Ovchinnikov}}, \bibinfo {author} {\bibfnamefont {D.}~\bibnamefont {Pasquier}}, \bibinfo {author} {\bibfnamefont {O.~V.}\ \bibnamefont {Yazyev}}, \ and\ \bibinfo {author} {\bibfnamefont {A.}~\bibnamefont {Kis}},\ }\href {\doibase 10.1038/natrevmats.2017.33} {\bibfield  {journal} {\bibinfo  {journal} {Nature Reviews Materials}\ }\textbf {\bibinfo {volume} {2}},\ \bibinfo {pages} {17033} (\bibinfo {year} {2017})}\BibitemShut {NoStop}%
\bibitem [{\citenamefont {Zhu}\ \emph {et~al.}(2011)\citenamefont {Zhu}, \citenamefont {Cheng},\ and\ \citenamefont {Schwingenschlögl}}]{zhu2011}%
  \BibitemOpen
  \bibfield  {author} {\bibinfo {author} {\bibfnamefont {Z.~Y.}\ \bibnamefont {Zhu}}, \bibinfo {author} {\bibfnamefont {Y.~C.}\ \bibnamefont {Cheng}}, \ and\ \bibinfo {author} {\bibfnamefont {U.}~\bibnamefont {Schwingenschlögl}},\ }\href {\doibase 10.1103/PhysRevB.84.153402} {\bibfield  {journal} {\bibinfo  {journal} {Physical Review B}\ }\textbf {\bibinfo {volume} {84}},\ \bibinfo {pages} {153402} (\bibinfo {year} {2011})}\BibitemShut {NoStop}%
\bibitem [{\citenamefont {Zhao}\ \emph {et~al.}(2013)\citenamefont {Zhao}, \citenamefont {Ghorannevis}, \citenamefont {Chu}, \citenamefont {Toh}, \citenamefont {Kloc}, \citenamefont {Tan},\ and\ \citenamefont {Eda}}]{zhao2013}%
  \BibitemOpen
  \bibfield  {author} {\bibinfo {author} {\bibfnamefont {W.}~\bibnamefont {Zhao}}, \bibinfo {author} {\bibfnamefont {Z.}~\bibnamefont {Ghorannevis}}, \bibinfo {author} {\bibfnamefont {L.}~\bibnamefont {Chu}}, \bibinfo {author} {\bibfnamefont {M.}~\bibnamefont {Toh}}, \bibinfo {author} {\bibfnamefont {C.}~\bibnamefont {Kloc}}, \bibinfo {author} {\bibfnamefont {P.~H.}\ \bibnamefont {Tan}}, \ and\ \bibinfo {author} {\bibfnamefont {G.}~\bibnamefont {Eda}},\ }\href {\doibase 10.1021/nn305275h} {\bibfield  {journal} {\bibinfo  {journal} {ACS Nano}\ }\textbf {\bibinfo {volume} {7}},\ \bibinfo {pages} {791} (\bibinfo {year} {2013})}\BibitemShut {NoStop}%
\bibitem [{\citenamefont {Ali}\ \emph {et~al.}(2014)\citenamefont {Ali}, \citenamefont {Xiong}, \citenamefont {Flynn}, \citenamefont {Tao}, \citenamefont {Gibson}, \citenamefont {Schoop}, \citenamefont {Liang}, \citenamefont {Haldolaarachchige}, \citenamefont {Hirschberger}, \citenamefont {Ong},\ and\ \citenamefont {Cava}}]{ali2014}%
  \BibitemOpen
  \bibfield  {author} {\bibinfo {author} {\bibfnamefont {M.~N.}\ \bibnamefont {Ali}}, \bibinfo {author} {\bibfnamefont {J.}~\bibnamefont {Xiong}}, \bibinfo {author} {\bibfnamefont {S.}~\bibnamefont {Flynn}}, \bibinfo {author} {\bibfnamefont {J.}~\bibnamefont {Tao}}, \bibinfo {author} {\bibfnamefont {Q.~D.}\ \bibnamefont {Gibson}}, \bibinfo {author} {\bibfnamefont {L.~M.}\ \bibnamefont {Schoop}}, \bibinfo {author} {\bibfnamefont {T.}~\bibnamefont {Liang}}, \bibinfo {author} {\bibfnamefont {N.}~\bibnamefont {Haldolaarachchige}}, \bibinfo {author} {\bibfnamefont {M.}~\bibnamefont {Hirschberger}}, \bibinfo {author} {\bibfnamefont {N.~P.}\ \bibnamefont {Ong}}, \ and\ \bibinfo {author} {\bibfnamefont {R.~J.}\ \bibnamefont {Cava}},\ }\href {\doibase 10.1038/nature13763} {\bibfield  {journal} {\bibinfo  {journal} {Nature}\ }\textbf {\bibinfo {volume} {514}},\ \bibinfo {pages} {205} (\bibinfo {year} {2014})}\BibitemShut {NoStop}%
\bibitem [{\citenamefont {Qi}\ \emph {et~al.}(2016)\citenamefont {Qi}, \citenamefont {Naumov}, \citenamefont {Ali}, \citenamefont {Rajamathi}, \citenamefont {Schnelle}, \citenamefont {Barkalov}, \citenamefont {Hanfland}, \citenamefont {Wu}, \citenamefont {Shekhar}, \citenamefont {Sun}, \citenamefont {Schmidt}, \citenamefont {Schwarz}, \citenamefont {Pippel}, \citenamefont {Werner}, \citenamefont {Hillebrand}, \citenamefont {Förster}, \citenamefont {Kampert}, \citenamefont {Parkin}, \citenamefont {Cava}, \citenamefont {Felser}, \citenamefont {Yan},\ and\ \citenamefont {Medvedev}}]{qi2016}%
  \BibitemOpen
  \bibfield  {author} {\bibinfo {author} {\bibfnamefont {Y.}~\bibnamefont {Qi}}, \bibinfo {author} {\bibfnamefont {P.~G.}\ \bibnamefont {Naumov}}, \bibinfo {author} {\bibfnamefont {M.~N.}\ \bibnamefont {Ali}}, \bibinfo {author} {\bibfnamefont {C.~R.}\ \bibnamefont {Rajamathi}}, \bibinfo {author} {\bibfnamefont {W.}~\bibnamefont {Schnelle}}, \bibinfo {author} {\bibfnamefont {O.}~\bibnamefont {Barkalov}}, \bibinfo {author} {\bibfnamefont {M.}~\bibnamefont {Hanfland}}, \bibinfo {author} {\bibfnamefont {S.-C.}\ \bibnamefont {Wu}}, \bibinfo {author} {\bibfnamefont {C.}~\bibnamefont {Shekhar}}, \bibinfo {author} {\bibfnamefont {Y.}~\bibnamefont {Sun}}, \bibinfo {author} {\bibfnamefont {M.}~\bibnamefont {Schmidt}}, \bibinfo {author} {\bibfnamefont {U.}~\bibnamefont {Schwarz}}, \bibinfo {author} {\bibfnamefont {E.}~\bibnamefont {Pippel}}, \bibinfo {author} {\bibfnamefont {P.}~\bibnamefont {Werner}}, \bibinfo {author} {\bibfnamefont {R.}~\bibnamefont {Hillebrand}}, \bibinfo {author} {\bibfnamefont {T.}~\bibnamefont
  {Förster}}, \bibinfo {author} {\bibfnamefont {E.}~\bibnamefont {Kampert}}, \bibinfo {author} {\bibfnamefont {S.~S.~P.}\ \bibnamefont {Parkin}}, \bibinfo {author} {\bibfnamefont {R.~J.}\ \bibnamefont {Cava}}, \bibinfo {author} {\bibfnamefont {C.}~\bibnamefont {Felser}}, \bibinfo {author} {\bibfnamefont {B.}~\bibnamefont {Yan}}, \ and\ \bibinfo {author} {\bibfnamefont {S.~A.}\ \bibnamefont {Medvedev}},\ }\href {\doibase 10.1038/ncomms11038} {\bibfield  {journal} {\bibinfo  {journal} {Nature Communications}\ }\textbf {\bibinfo {volume} {7}},\ \bibinfo {pages} {11038} (\bibinfo {year} {2016})}\BibitemShut {NoStop}%
\bibitem [{\citenamefont {Deng}\ \emph {et~al.}(2016)\citenamefont {Deng}, \citenamefont {Wan}, \citenamefont {Deng}, \citenamefont {Zhang}, \citenamefont {Ding}, \citenamefont {Wang}, \citenamefont {Yan}, \citenamefont {Huang}, \citenamefont {Zhang}, \citenamefont {Xu}, \citenamefont {Denlinger}, \citenamefont {Fedorov}, \citenamefont {Yang}, \citenamefont {Duan}, \citenamefont {Yao}, \citenamefont {Wu}, \citenamefont {Fan}, \citenamefont {Zhang},\ and\ \citenamefont {Zhou}}]{deng2016}%
  \BibitemOpen
  \bibfield  {author} {\bibinfo {author} {\bibfnamefont {K.}~\bibnamefont {Deng}}, \bibinfo {author} {\bibfnamefont {G.}~\bibnamefont {Wan}}, \bibinfo {author} {\bibfnamefont {P.}~\bibnamefont {Deng}}, \bibinfo {author} {\bibfnamefont {K.}~\bibnamefont {Zhang}}, \bibinfo {author} {\bibfnamefont {S.}~\bibnamefont {Ding}}, \bibinfo {author} {\bibfnamefont {E.}~\bibnamefont {Wang}}, \bibinfo {author} {\bibfnamefont {M.}~\bibnamefont {Yan}}, \bibinfo {author} {\bibfnamefont {H.}~\bibnamefont {Huang}}, \bibinfo {author} {\bibfnamefont {H.}~\bibnamefont {Zhang}}, \bibinfo {author} {\bibfnamefont {Z.}~\bibnamefont {Xu}}, \bibinfo {author} {\bibfnamefont {J.}~\bibnamefont {Denlinger}}, \bibinfo {author} {\bibfnamefont {A.}~\bibnamefont {Fedorov}}, \bibinfo {author} {\bibfnamefont {H.}~\bibnamefont {Yang}}, \bibinfo {author} {\bibfnamefont {W.}~\bibnamefont {Duan}}, \bibinfo {author} {\bibfnamefont {H.}~\bibnamefont {Yao}}, \bibinfo {author} {\bibfnamefont {Y.}~\bibnamefont {Wu}}, \bibinfo {author} {\bibfnamefont
  {S.}~\bibnamefont {Fan}}, \bibinfo {author} {\bibfnamefont {A.}~\bibnamefont {Zhang}}, \ and\ \bibinfo {author} {\bibfnamefont {S.}~\bibnamefont {Zhou}},\ }\href {\doibase 10.1038/nphys3871} {\bibfield  {journal} {\bibinfo  {journal} {Nature Physics}\ }\textbf {\bibinfo {volume} {12}},\ \bibinfo {pages} {1105} (\bibinfo {year} {2016})}\BibitemShut {NoStop}%
\bibitem [{\citenamefont {Wang}\ \emph {et~al.}(2016)\citenamefont {Wang}, \citenamefont {Zhou}, \citenamefont {Chien},\ and\ \citenamefont {Wray}}]{wang2016}%
  \BibitemOpen
  \bibfield  {author} {\bibinfo {author} {\bibfnamefont {Y.}~\bibnamefont {Wang}}, \bibinfo {author} {\bibfnamefont {Y.}~\bibnamefont {Zhou}}, \bibinfo {author} {\bibfnamefont {C.-L.}\ \bibnamefont {Chien}}, \ and\ \bibinfo {author} {\bibfnamefont {L.~A.}\ \bibnamefont {Wray}},\ }\href {\doibase 10.1103/PhysRevB.93.121108} {\bibfield  {journal} {\bibinfo  {journal} {Phys. Rev. B}\ }\textbf {\bibinfo {volume} {93}},\ \bibinfo {pages} {121108} (\bibinfo {year} {2016})}\BibitemShut {NoStop}%
\bibitem [{\citenamefont {Paudyal}\ \emph {et~al.}(2020)\citenamefont {Paudyal}, \citenamefont {Ponc{\'e}}, \citenamefont {Giustino},\ and\ \citenamefont {Margine}}]{paudyal2020}%
  \BibitemOpen
  \bibfield  {author} {\bibinfo {author} {\bibfnamefont {H.}~\bibnamefont {Paudyal}}, \bibinfo {author} {\bibfnamefont {S.}~\bibnamefont {Ponc{\'e}}}, \bibinfo {author} {\bibfnamefont {F.}~\bibnamefont {Giustino}}, \ and\ \bibinfo {author} {\bibfnamefont {E.~R.}\ \bibnamefont {Margine}},\ }\href {https://journals.aps.org/prb/pdf/10.1103/PhysRevB.101.214515} {\bibfield  {journal} {\bibinfo  {journal} {Physical Review B}\ }\textbf {\bibinfo {volume} {101}},\ \bibinfo {pages} {214515} (\bibinfo {year} {2020})}\BibitemShut {NoStop}%
\bibitem [{\citenamefont {Morosan}\ \emph {et~al.}(2006)\citenamefont {Morosan}, \citenamefont {Zandbergen}, \citenamefont {Dennis}, \citenamefont {Bos}, \citenamefont {Onose}, \citenamefont {Klimczuk}, \citenamefont {Ramirez}, \citenamefont {Ong},\ and\ \citenamefont {Cava}}]{morosan2006}%
  \BibitemOpen
  \bibfield  {author} {\bibinfo {author} {\bibfnamefont {E.}~\bibnamefont {Morosan}}, \bibinfo {author} {\bibfnamefont {H.~W.}\ \bibnamefont {Zandbergen}}, \bibinfo {author} {\bibfnamefont {B.~S.}\ \bibnamefont {Dennis}}, \bibinfo {author} {\bibfnamefont {J.~W.~G.}\ \bibnamefont {Bos}}, \bibinfo {author} {\bibfnamefont {Y.}~\bibnamefont {Onose}}, \bibinfo {author} {\bibfnamefont {T.}~\bibnamefont {Klimczuk}}, \bibinfo {author} {\bibfnamefont {A.~P.}\ \bibnamefont {Ramirez}}, \bibinfo {author} {\bibfnamefont {N.~P.}\ \bibnamefont {Ong}}, \ and\ \bibinfo {author} {\bibfnamefont {R.~J.}\ \bibnamefont {Cava}},\ }\href {\doibase 10.1038/nphys360} {\bibfield  {journal} {\bibinfo  {journal} {Nature Physics}\ }\textbf {\bibinfo {volume} {2}},\ \bibinfo {pages} {544} (\bibinfo {year} {2006})}\BibitemShut {NoStop}%
\bibitem [{\citenamefont {Rossnagel}(2011)}]{rossnagel2011}%
  \BibitemOpen
  \bibfield  {author} {\bibinfo {author} {\bibfnamefont {K.}~\bibnamefont {Rossnagel}},\ }\href {\doibase 10.1088/0953-8984/23/21/213001} {\bibfield  {journal} {\bibinfo  {journal} {Journal of Physics: Condensed Matter}\ }\textbf {\bibinfo {volume} {23}},\ \bibinfo {pages} {213001} (\bibinfo {year} {2011})}\BibitemShut {NoStop}%
\bibitem [{\citenamefont {Soumyanarayanan}\ \emph {et~al.}(2013)\citenamefont {Soumyanarayanan}, \citenamefont {Yee}, \citenamefont {He}, \citenamefont {van Wezel}, \citenamefont {Rahn}, \citenamefont {Rossnagel}, \citenamefont {Hudson}, \citenamefont {Norman},\ and\ \citenamefont {Hoffman}}]{soumyanarayanan2013}%
  \BibitemOpen
  \bibfield  {author} {\bibinfo {author} {\bibfnamefont {A.}~\bibnamefont {Soumyanarayanan}}, \bibinfo {author} {\bibfnamefont {M.~M.}\ \bibnamefont {Yee}}, \bibinfo {author} {\bibfnamefont {Y.}~\bibnamefont {He}}, \bibinfo {author} {\bibfnamefont {J.}~\bibnamefont {van Wezel}}, \bibinfo {author} {\bibfnamefont {D.~J.}\ \bibnamefont {Rahn}}, \bibinfo {author} {\bibfnamefont {K.}~\bibnamefont {Rossnagel}}, \bibinfo {author} {\bibfnamefont {E.~W.}\ \bibnamefont {Hudson}}, \bibinfo {author} {\bibfnamefont {M.~R.}\ \bibnamefont {Norman}}, \ and\ \bibinfo {author} {\bibfnamefont {J.~E.}\ \bibnamefont {Hoffman}},\ }\href {\doibase 10.1073/pnas.1211387110} {\bibfield  {journal} {\bibinfo  {journal} {Proceedings of the National Academy of Sciences}\ }\textbf {\bibinfo {volume} {110}},\ \bibinfo {pages} {1623} (\bibinfo {year} {2013})}\BibitemShut {NoStop}%
\bibitem [{\citenamefont {Xi}\ \emph {et~al.}(2015)\citenamefont {Xi}, \citenamefont {Wang}, \citenamefont {Zhao}, \citenamefont {Park}, \citenamefont {Law}, \citenamefont {Berger}, \citenamefont {Forró}, \citenamefont {Shan},\ and\ \citenamefont {Mak}}]{xi2015cdw}%
  \BibitemOpen
  \bibfield  {author} {\bibinfo {author} {\bibfnamefont {X.}~\bibnamefont {Xi}}, \bibinfo {author} {\bibfnamefont {L.}~\bibnamefont {Wang}}, \bibinfo {author} {\bibfnamefont {W.}~\bibnamefont {Zhao}}, \bibinfo {author} {\bibfnamefont {J.-H.}\ \bibnamefont {Park}}, \bibinfo {author} {\bibfnamefont {K.~T.}\ \bibnamefont {Law}}, \bibinfo {author} {\bibfnamefont {H.}~\bibnamefont {Berger}}, \bibinfo {author} {\bibfnamefont {L.}~\bibnamefont {Forró}}, \bibinfo {author} {\bibfnamefont {J.}~\bibnamefont {Shan}}, \ and\ \bibinfo {author} {\bibfnamefont {K.~F.}\ \bibnamefont {Mak}},\ }\href {\doibase 10.1038/nnano.2015.143} {\bibfield  {journal} {\bibinfo  {journal} {Nature Nanotechnology}\ }\textbf {\bibinfo {volume} {10}},\ \bibinfo {pages} {765} (\bibinfo {year} {2015})}\BibitemShut {NoStop}%
\bibitem [{\citenamefont {Xi}\ \emph {et~al.}(2016)\citenamefont {Xi}, \citenamefont {Zhao}, \citenamefont {Wang}, \citenamefont {Berger}, \citenamefont {Forró}, \citenamefont {Shan},\ and\ \citenamefont {Mak}}]{xi2016ising}%
  \BibitemOpen
  \bibfield  {author} {\bibinfo {author} {\bibfnamefont {X.}~\bibnamefont {Xi}}, \bibinfo {author} {\bibfnamefont {W.}~\bibnamefont {Zhao}}, \bibinfo {author} {\bibfnamefont {Z.}~\bibnamefont {Wang}}, \bibinfo {author} {\bibfnamefont {H.}~\bibnamefont {Berger}}, \bibinfo {author} {\bibfnamefont {L.}~\bibnamefont {Forró}}, \bibinfo {author} {\bibfnamefont {J.}~\bibnamefont {Shan}}, \ and\ \bibinfo {author} {\bibfnamefont {K.~F.}\ \bibnamefont {Mak}},\ }\href {\doibase 10.1038/nphys3538} {\bibfield  {journal} {\bibinfo  {journal} {Nature Physics}\ }\textbf {\bibinfo {volume} {12}},\ \bibinfo {pages} {139–143} (\bibinfo {year} {2016})}\BibitemShut {NoStop}%
\bibitem [{\citenamefont {Yokoya}\ \emph {et~al.}(2001)\citenamefont {Yokoya}, \citenamefont {Kiss}, \citenamefont {Chainani}, \citenamefont {Shin}, \citenamefont {Nohara},\ and\ \citenamefont {Takagi}}]{yokoya2001}%
  \BibitemOpen
  \bibfield  {author} {\bibinfo {author} {\bibfnamefont {T.}~\bibnamefont {Yokoya}}, \bibinfo {author} {\bibfnamefont {T.}~\bibnamefont {Kiss}}, \bibinfo {author} {\bibfnamefont {A.}~\bibnamefont {Chainani}}, \bibinfo {author} {\bibfnamefont {S.}~\bibnamefont {Shin}}, \bibinfo {author} {\bibfnamefont {M.}~\bibnamefont {Nohara}}, \ and\ \bibinfo {author} {\bibfnamefont {H.}~\bibnamefont {Takagi}},\ }\href {\doibase 10.1126/science.1065067} {\bibfield  {journal} {\bibinfo  {journal} {Science}\ }\textbf {\bibinfo {volume} {294}},\ \bibinfo {pages} {2518} (\bibinfo {year} {2001})}\BibitemShut {NoStop}%
\bibitem [{\citenamefont {Kiss}\ \emph {et~al.}(2007)\citenamefont {Kiss}, \citenamefont {Yokoya}, \citenamefont {Chainani}, \citenamefont {Shin}, \citenamefont {Hanaguri}, \citenamefont {Nohara},\ and\ \citenamefont {Takagi}}]{kiss2007}%
  \BibitemOpen
  \bibfield  {author} {\bibinfo {author} {\bibfnamefont {T.}~\bibnamefont {Kiss}}, \bibinfo {author} {\bibfnamefont {T.}~\bibnamefont {Yokoya}}, \bibinfo {author} {\bibfnamefont {A.}~\bibnamefont {Chainani}}, \bibinfo {author} {\bibfnamefont {S.}~\bibnamefont {Shin}}, \bibinfo {author} {\bibfnamefont {T.}~\bibnamefont {Hanaguri}}, \bibinfo {author} {\bibfnamefont {M.}~\bibnamefont {Nohara}}, \ and\ \bibinfo {author} {\bibfnamefont {H.}~\bibnamefont {Takagi}},\ }\href {\doibase 10.1038/nphys699} {\bibfield  {journal} {\bibinfo  {journal} {Nature Physics}\ }\textbf {\bibinfo {volume} {3}},\ \bibinfo {pages} {720} (\bibinfo {year} {2007})}\BibitemShut {NoStop}%
\bibitem [{\citenamefont {Borisenko}\ \emph {et~al.}(2009)\citenamefont {Borisenko}, \citenamefont {Kordyuk}, \citenamefont {Zabolotnyy}, \citenamefont {Evtushinsky}, \citenamefont {Kim}, \citenamefont {Morozov}, \citenamefont {Yaresko}, \citenamefont {Vasiliev}, \citenamefont {Follath},\ and\ \citenamefont {Büchner}}]{borisenko2009}%
  \BibitemOpen
  \bibfield  {author} {\bibinfo {author} {\bibfnamefont {S.~V.}\ \bibnamefont {Borisenko}}, \bibinfo {author} {\bibfnamefont {A.~A.}\ \bibnamefont {Kordyuk}}, \bibinfo {author} {\bibfnamefont {V.~B.}\ \bibnamefont {Zabolotnyy}}, \bibinfo {author} {\bibfnamefont {D.~V.}\ \bibnamefont {Evtushinsky}}, \bibinfo {author} {\bibfnamefont {T.~K.}\ \bibnamefont {Kim}}, \bibinfo {author} {\bibfnamefont {I.~V.}\ \bibnamefont {Morozov}}, \bibinfo {author} {\bibfnamefont {A.~N.}\ \bibnamefont {Yaresko}}, \bibinfo {author} {\bibfnamefont {A.~N.}\ \bibnamefont {Vasiliev}}, \bibinfo {author} {\bibfnamefont {R.}~\bibnamefont {Follath}}, \ and\ \bibinfo {author} {\bibfnamefont {B.}~\bibnamefont {Büchner}},\ }\href {\doibase 10.1103/PhysRevLett.102.166402} {\bibfield  {journal} {\bibinfo  {journal} {Physical Review Letters}\ }\textbf {\bibinfo {volume} {102}},\ \bibinfo {pages} {166402} (\bibinfo {year} {2009})}\BibitemShut {NoStop}%
\bibitem [{\citenamefont {Das}\ \emph {et~al.}(2023)\citenamefont {Das}, \citenamefont {Paudyal}, \citenamefont {Margine}, \citenamefont {Agterberg},\ and\ \citenamefont {Mazin}}]{Das2022s}%
  \BibitemOpen
  \bibfield  {author} {\bibinfo {author} {\bibfnamefont {S.}~\bibnamefont {Das}}, \bibinfo {author} {\bibfnamefont {H.}~\bibnamefont {Paudyal}}, \bibinfo {author} {\bibfnamefont {E.}~\bibnamefont {Margine}}, \bibinfo {author} {\bibfnamefont {D.}~\bibnamefont {Agterberg}}, \ and\ \bibinfo {author} {\bibfnamefont {I.}~\bibnamefont {Mazin}},\ }\href {\doibase 10.1038/s41524-023-01017-4} {\bibfield  {journal} {\bibinfo  {journal} {npj Comput. Mater.}\ }\textbf {\bibinfo {volume} {9}},\ \bibinfo {pages} {66} (\bibinfo {year} {2023})}\BibitemShut {NoStop}%
\bibitem [{\citenamefont {Nakata}\ \emph {et~al.}(2018)\citenamefont {Nakata}, \citenamefont {Sugawara}, \citenamefont {Ichinokura}, \citenamefont {Okada}, \citenamefont {Hitosugi}, \citenamefont {Koretsune}, \citenamefont {Ueno}, \citenamefont {Hasegawa},\ and\ \citenamefont {Takahashi}}]{nakata2018anisotropic}%
  \BibitemOpen
  \bibfield  {author} {\bibinfo {author} {\bibfnamefont {Y.}~\bibnamefont {Nakata}}, \bibinfo {author} {\bibfnamefont {K.}~\bibnamefont {Sugawara}}, \bibinfo {author} {\bibfnamefont {S.}~\bibnamefont {Ichinokura}}, \bibinfo {author} {\bibfnamefont {Y.}~\bibnamefont {Okada}}, \bibinfo {author} {\bibfnamefont {T.}~\bibnamefont {Hitosugi}}, \bibinfo {author} {\bibfnamefont {T.}~\bibnamefont {Koretsune}}, \bibinfo {author} {\bibfnamefont {K.}~\bibnamefont {Ueno}}, \bibinfo {author} {\bibfnamefont {S.}~\bibnamefont {Hasegawa}}, \ and\ \bibinfo {author} {\bibfnamefont {T.}~\bibnamefont {Takahashi}},\ }\href {\doibase 10.1038/s41699-018-0057-3} {\bibfield  {journal} {\bibinfo  {journal} {npj 2D Materials and Applications}\ }\textbf {\bibinfo {volume} {2}},\ \bibinfo {pages} {12} (\bibinfo {year} {2018})}\BibitemShut {NoStop}%
\bibitem [{\citenamefont {Kundu}\ \emph {et~al.}(2024)\citenamefont {Kundu}, \citenamefont {Rajapitamahuni}, \citenamefont {Vescovo}, \citenamefont {Klimovskikh}, \citenamefont {Berger},\ and\ \citenamefont {Valla}}]{kundu2024cdw}%
  \BibitemOpen
  \bibfield  {author} {\bibinfo {author} {\bibfnamefont {A.~K.}\ \bibnamefont {Kundu}}, \bibinfo {author} {\bibfnamefont {A.}~\bibnamefont {Rajapitamahuni}}, \bibinfo {author} {\bibfnamefont {E.}~\bibnamefont {Vescovo}}, \bibinfo {author} {\bibfnamefont {I.~I.}\ \bibnamefont {Klimovskikh}}, \bibinfo {author} {\bibfnamefont {H.}~\bibnamefont {Berger}}, \ and\ \bibinfo {author} {\bibfnamefont {T.}~\bibnamefont {Valla}},\ }\href {\doibase 10.1038/s43246-024-00661-7} {\bibfield  {journal} {\bibinfo  {journal} {Communications Materials}\ }\textbf {\bibinfo {volume} {5}},\ \bibinfo {pages} {208} (\bibinfo {year} {2024})}\BibitemShut {NoStop}%
\bibitem [{\citenamefont {Weber}\ \emph {et~al.}(2011)\citenamefont {Weber}, \citenamefont {Rosenkranz}, \citenamefont {Castellan}, \citenamefont {Osborn}, \citenamefont {Hott}, \citenamefont {Heid}, \citenamefont {Bohnen}, \citenamefont {Egami}, \citenamefont {Said},\ and\ \citenamefont {Reznik}}]{Weber2011s}%
  \BibitemOpen
  \bibfield  {author} {\bibinfo {author} {\bibfnamefont {F.}~\bibnamefont {Weber}}, \bibinfo {author} {\bibfnamefont {S.}~\bibnamefont {Rosenkranz}}, \bibinfo {author} {\bibfnamefont {J.~P.}\ \bibnamefont {Castellan}}, \bibinfo {author} {\bibfnamefont {R.}~\bibnamefont {Osborn}}, \bibinfo {author} {\bibfnamefont {R.}~\bibnamefont {Hott}}, \bibinfo {author} {\bibfnamefont {R.}~\bibnamefont {Heid}}, \bibinfo {author} {\bibfnamefont {K.-P.}\ \bibnamefont {Bohnen}}, \bibinfo {author} {\bibfnamefont {T.}~\bibnamefont {Egami}}, \bibinfo {author} {\bibfnamefont {A.~H.}\ \bibnamefont {Said}}, \ and\ \bibinfo {author} {\bibfnamefont {D.}~\bibnamefont {Reznik}},\ }\href {\doibase 10.1103/PhysRevLett.107.107403} {\bibfield  {journal} {\bibinfo  {journal} {Phys. Rev. Lett.}\ }\textbf {\bibinfo {volume} {107}},\ \bibinfo {pages} {107403} (\bibinfo {year} {2011})}\BibitemShut {NoStop}%
\bibitem [{\citenamefont {Zheng}\ and\ \citenamefont {Feng}(2019)}]{zheng2019s}%
  \BibitemOpen
  \bibfield  {author} {\bibinfo {author} {\bibfnamefont {F.}~\bibnamefont {Zheng}}\ and\ \bibinfo {author} {\bibfnamefont {J.}~\bibnamefont {Feng}},\ }\href {\doibase 10.1103/PhysRevB.99.161119} {\bibfield  {journal} {\bibinfo  {journal} {Phys. Rev. B}\ }\textbf {\bibinfo {volume} {99}},\ \bibinfo {pages} {161119} (\bibinfo {year} {2019})}\BibitemShut {NoStop}%
\bibitem [{\citenamefont {Suderow}\ \emph {et~al.}(2005)\citenamefont {Suderow}, \citenamefont {Tissen}, \citenamefont {Brison}, \citenamefont {Martínez},\ and\ \citenamefont {Vieira}}]{suderow2005}%
  \BibitemOpen
  \bibfield  {author} {\bibinfo {author} {\bibfnamefont {H.}~\bibnamefont {Suderow}}, \bibinfo {author} {\bibfnamefont {V.~G.}\ \bibnamefont {Tissen}}, \bibinfo {author} {\bibfnamefont {J.~P.}\ \bibnamefont {Brison}}, \bibinfo {author} {\bibfnamefont {J.~L.}\ \bibnamefont {Martínez}}, \ and\ \bibinfo {author} {\bibfnamefont {S.}~\bibnamefont {Vieira}},\ }\href {\doibase 10.1088/0953-2048/18/6/010} {\bibfield  {journal} {\bibinfo  {journal} {Superconductor Science and Technology}\ }\textbf {\bibinfo {volume} {18}},\ \bibinfo {pages} {598} (\bibinfo {year} {2005})}\BibitemShut {NoStop}%
\bibitem [{\citenamefont {Calandra}(2011)}]{calandra2011}%
  \BibitemOpen
  \bibfield  {author} {\bibinfo {author} {\bibfnamefont {M.}~\bibnamefont {Calandra}},\ }\href {\doibase 10.1088/0953-8984/23/10/102201} {\bibfield  {journal} {\bibinfo  {journal} {Journal of Physics: Condensed Matter}\ }\textbf {\bibinfo {volume} {23}},\ \bibinfo {pages} {102201} (\bibinfo {year} {2011})}\BibitemShut {NoStop}%
\bibitem [{\citenamefont {Morosan}\ \emph {et~al.}(1991)\citenamefont {Morosan}, \citenamefont {Zandbergen}, \citenamefont {Dennis}, \citenamefont {Bos}, \citenamefont {Onose}, \citenamefont {Klimczuk}, \citenamefont {Ramirez}, \citenamefont {Ong},\ and\ \citenamefont {Cava}}]{morosan1991}%
  \BibitemOpen
  \bibfield  {author} {\bibinfo {author} {\bibfnamefont {E.}~\bibnamefont {Morosan}}, \bibinfo {author} {\bibfnamefont {H.~W.}\ \bibnamefont {Zandbergen}}, \bibinfo {author} {\bibfnamefont {B.~S.}\ \bibnamefont {Dennis}}, \bibinfo {author} {\bibfnamefont {J.~W.~G.}\ \bibnamefont {Bos}}, \bibinfo {author} {\bibfnamefont {Y.}~\bibnamefont {Onose}}, \bibinfo {author} {\bibfnamefont {T.}~\bibnamefont {Klimczuk}}, \bibinfo {author} {\bibfnamefont {A.~P.}\ \bibnamefont {Ramirez}}, \bibinfo {author} {\bibfnamefont {N.~P.}\ \bibnamefont {Ong}}, \ and\ \bibinfo {author} {\bibfnamefont {R.~J.}\ \bibnamefont {Cava}},\ }\href {\doibase 10.1016/0379-6779(91)91983-H} {\bibfield  {journal} {\bibinfo  {journal} {Chemistry of Materials}\ }\textbf {\bibinfo {volume} {3}},\ \bibinfo {pages} {493} (\bibinfo {year} {1991})}\BibitemShut {NoStop}%
\bibitem [{QE_()}]{QE_methodology}%
  \BibitemOpen
  \href@noop {} {}\bibinfo {note} {\textit{Ab initio} calculations are performed using \texttt{Quantum ESPRESSO}~\cite{giannozzi2017o} package with relativistic norm-conserving pseudopotentials~\cite{PseudoDojo2018} within the Perdew-Burke-Ernzerhof~\cite{perdew1996o} exchange-correlation functional in the generalized gradient approximation. A plane wave kinetic-energy (charge density) cutoff value of 80~Ry (320~Ry), a $\Gamma$-centered $8\times 8 \times 1$ Monkhorst-Pack \textbf{k}-mesh~\cite{Monkhors1976s}, and a Methfessel and Paxton smearing~\cite{Methfessel1989s} width of 0.01~Ry have been used for the Brillouin-zone sampling. The atomic positions and lattice parameters are optimized until the self-consistent energy is converged within $2.7\times10^{-5}$~eV and the maximum force on each atom is less than 0.005~eV/\AA. For the DOS and Fermi surface calculations, denser \textbf{k}-meshes of $16 \times 16 \times 16$ and $32 \times 32 \times 32$ are used. The dynamical matrices and the linear variation of
  the self-consistent potential are calculated within density-functional perturbation theory~\cite{Baroni2001s} on an irreducible $2 \times 2 \times 1$ \textbf{q} mesh.}\BibitemShut {Stop}%
\bibitem [{EPW()}]{EPW_methodology}%
  \BibitemOpen
  \href@noop {} {}\bibinfo {note} {Migdal-Eliashberg theory~\cite{Allen1975s, Margine2013s}, as implemented in the Electron-Phonon Wannier (\texttt{EPW}) code~\cite{EPW2023}, is used to investigate the electron-phonon and superconducting properties. Denser (uniform) $20\times 20 \times 20$ \textbf{k}- and \textbf{q}-point grids are used for electron-phonon matrix element calculations. Effective Coulomb repulsion value of $\mu^* = 0.2$ is used to estimate the superconducting critical temperature by self-consistently solving the isotropic/anisotropic Migdal-Eliashberg equations. The Matsubara frequency cutoff is set to 1.5~eV and the Dirac deltas are replaced by Lorentzians of width 50~meV for electrons and 0.1~meV for phonons.}\BibitemShut {Stop}%
\bibitem [{W90()}]{W90_methodology}%
  \BibitemOpen
  \href@noop {} {}\bibinfo {note} {Wannier interpolation~\cite{Pizzi2019s} on a uniform $8 \times 8 \times 8$ $\Gamma$-centered \textbf{k}-grid is used to calculate the Nb $4d$ and Se $3p$ orbital Wannier functions used in the electron-phonon calculations.}\BibitemShut {Stop}%
\bibitem [{\citenamefont {Wilson}\ \emph {et~al.}(1975)\citenamefont {Wilson}, \citenamefont {Di~Salvo},\ and\ \citenamefont {Mahajan}}]{wilson1975}%
  \BibitemOpen
  \bibfield  {author} {\bibinfo {author} {\bibfnamefont {J.~A.}\ \bibnamefont {Wilson}}, \bibinfo {author} {\bibfnamefont {F.~J.}\ \bibnamefont {Di~Salvo}}, \ and\ \bibinfo {author} {\bibfnamefont {S.}~\bibnamefont {Mahajan}},\ }\href {\doibase 10.1080/00018737500101391} {\bibfield  {journal} {\bibinfo  {journal} {Advances in Physics}\ }\textbf {\bibinfo {volume} {24}},\ \bibinfo {pages} {117} (\bibinfo {year} {1975})}\BibitemShut {NoStop}%
\bibitem [{\citenamefont {Novoselov}\ \emph {et~al.}(2005)\citenamefont {Novoselov}, \citenamefont {Jiang}, \citenamefont {Schedin}, \citenamefont {Booth}, \citenamefont {Khotkevich}, \citenamefont {Morozov},\ and\ \citenamefont {Geim}}]{novoselov2005mechanical}%
  \BibitemOpen
  \bibfield  {author} {\bibinfo {author} {\bibfnamefont {K.~S.}\ \bibnamefont {Novoselov}}, \bibinfo {author} {\bibfnamefont {D.}~\bibnamefont {Jiang}}, \bibinfo {author} {\bibfnamefont {F.}~\bibnamefont {Schedin}}, \bibinfo {author} {\bibfnamefont {T.~J.}\ \bibnamefont {Booth}}, \bibinfo {author} {\bibfnamefont {V.~V.}\ \bibnamefont {Khotkevich}}, \bibinfo {author} {\bibfnamefont {S.~V.}\ \bibnamefont {Morozov}}, \ and\ \bibinfo {author} {\bibfnamefont {A.~K.}\ \bibnamefont {Geim}},\ }\href {10.1073/pnas.0502848102} {\bibfield  {journal} {\bibinfo  {journal} {PNAS}\ }\textbf {\bibinfo {volume} {102}},\ \bibinfo {pages} {10451} (\bibinfo {year} {2005})}\BibitemShut {NoStop}%
\bibitem [{\citenamefont {Yoshizawa}\ \emph {et~al.}(2024)\citenamefont {Yoshizawa}, \citenamefont {Sagisaka},\ and\ \citenamefont {Sakata}}]{Yoshizawa2024}%
  \BibitemOpen
  \bibfield  {author} {\bibinfo {author} {\bibfnamefont {S.}~\bibnamefont {Yoshizawa}}, \bibinfo {author} {\bibfnamefont {K.}~\bibnamefont {Sagisaka}}, \ and\ \bibinfo {author} {\bibfnamefont {H.}~\bibnamefont {Sakata}},\ }\href {\doibase 10.1103/PhysRevLett.132.056401} {\bibfield  {journal} {\bibinfo  {journal} {Phys. Rev. Lett.}\ }\textbf {\bibinfo {volume} {132}},\ \bibinfo {pages} {056401} (\bibinfo {year} {2024})}\BibitemShut {NoStop}%
\bibitem [{\citenamefont {Sivakumar}\ \emph {et~al.}(2025)\citenamefont {Sivakumar}, \citenamefont {Aretz}, \citenamefont {Scherb}, \citenamefont {van Midden~Mavrič}, \citenamefont {Huijgen}, \citenamefont {Kamber}, \citenamefont {Wegner}, \citenamefont {Khajetoorians}, \citenamefont {Rösner},\ and\ \citenamefont {Hauptmann}}]{Sivakumar2025}%
  \BibitemOpen
  \bibfield  {author} {\bibinfo {author} {\bibfnamefont {N.~S.}\ \bibnamefont {Sivakumar}}, \bibinfo {author} {\bibfnamefont {J.}~\bibnamefont {Aretz}}, \bibinfo {author} {\bibfnamefont {S.}~\bibnamefont {Scherb}}, \bibinfo {author} {\bibfnamefont {M.}~\bibnamefont {van Midden~Mavrič}}, \bibinfo {author} {\bibfnamefont {N.}~\bibnamefont {Huijgen}}, \bibinfo {author} {\bibfnamefont {U.}~\bibnamefont {Kamber}}, \bibinfo {author} {\bibfnamefont {D.}~\bibnamefont {Wegner}}, \bibinfo {author} {\bibfnamefont {A.~A.}\ \bibnamefont {Khajetoorians}}, \bibinfo {author} {\bibfnamefont {M.}~\bibnamefont {Rösner}}, \ and\ \bibinfo {author} {\bibfnamefont {N.}~\bibnamefont {Hauptmann}},\ }\href {\doibase 10.1103/PhysRevB.111.075409} {\bibfield  {journal} {\bibinfo  {journal} {Phys. Rev. B}\ }\textbf {\bibinfo {volume} {111}},\ \bibinfo {pages} {075409} (\bibinfo {year} {2025})}\BibitemShut {NoStop}%
\bibitem [{\citenamefont {Zheng}\ \emph {et~al.}(2018)\citenamefont {Zheng}, \citenamefont {Zhou}, \citenamefont {Liu},\ and\ \citenamefont {Feng}}]{zheng2018}%
  \BibitemOpen
  \bibfield  {author} {\bibinfo {author} {\bibfnamefont {F.}~\bibnamefont {Zheng}}, \bibinfo {author} {\bibfnamefont {Z.}~\bibnamefont {Zhou}}, \bibinfo {author} {\bibfnamefont {X.}~\bibnamefont {Liu}}, \ and\ \bibinfo {author} {\bibfnamefont {J.}~\bibnamefont {Feng}},\ }\href {\doibase 10.1103/PhysRevB.97.081101} {\bibfield  {journal} {\bibinfo  {journal} {Physical Review B}\ }\textbf {\bibinfo {volume} {97}},\ \bibinfo {pages} {081101} (\bibinfo {year} {2018})}\BibitemShut {NoStop}%
\bibitem [{\citenamefont {Johannes}\ and\ \citenamefont {Mazin}(2008)}]{johannes2008fermi}%
  \BibitemOpen
  \bibfield  {author} {\bibinfo {author} {\bibfnamefont {M.~D.}\ \bibnamefont {Johannes}}\ and\ \bibinfo {author} {\bibfnamefont {I.~I.}\ \bibnamefont {Mazin}},\ }\href {\doibase 10.1103/PhysRevB.77.165135} {\bibfield  {journal} {\bibinfo  {journal} {Physical Review B}\ }\textbf {\bibinfo {volume} {77}},\ \bibinfo {pages} {165135} (\bibinfo {year} {2008})}\BibitemShut {NoStop}%
\bibitem [{\citenamefont {Calandra}\ and\ \citenamefont {Mauri}(2011)}]{calandra2011cdw}%
  \BibitemOpen
  \bibfield  {author} {\bibinfo {author} {\bibfnamefont {M.}~\bibnamefont {Calandra}}\ and\ \bibinfo {author} {\bibfnamefont {F.}~\bibnamefont {Mauri}},\ }\href {\doibase 10.1103/PhysRevLett.106.196406} {\bibfield  {journal} {\bibinfo  {journal} {Physical Review Letters}\ }\textbf {\bibinfo {volume} {106}},\ \bibinfo {pages} {196406} (\bibinfo {year} {2011})}\BibitemShut {NoStop}%
\bibitem [{\citenamefont {Ying}\ \emph {et~al.}(2018)\citenamefont {Ying}, \citenamefont {Paudyal}, \citenamefont {Heil}, \citenamefont {Chen}, \citenamefont {Struzhkin},\ and\ \citenamefont {Margine}}]{ying2018unusual}%
  \BibitemOpen
  \bibfield  {author} {\bibinfo {author} {\bibfnamefont {J.}~\bibnamefont {Ying}}, \bibinfo {author} {\bibfnamefont {H.}~\bibnamefont {Paudyal}}, \bibinfo {author} {\bibfnamefont {C.}~\bibnamefont {Heil}}, \bibinfo {author} {\bibfnamefont {X.-J.}\ \bibnamefont {Chen}}, \bibinfo {author} {\bibfnamefont {V.~V.}\ \bibnamefont {Struzhkin}}, \ and\ \bibinfo {author} {\bibfnamefont {E.~R.}\ \bibnamefont {Margine}},\ }\href {\doibase 10.1103/PhysRevLett.121.027003} {\bibfield  {journal} {\bibinfo  {journal} {Physical Review Letters}\ }\textbf {\bibinfo {volume} {121}},\ \bibinfo {pages} {027003} (\bibinfo {year} {2018})}\BibitemShut {NoStop}%
\bibitem [{\citenamefont {Lian}\ \emph {et~al.}(2018)\citenamefont {Lian}, \citenamefont {Si},\ and\ \citenamefont {Duan}}]{lian2018}%
  \BibitemOpen
  \bibfield  {author} {\bibinfo {author} {\bibfnamefont {C.-S.}\ \bibnamefont {Lian}}, \bibinfo {author} {\bibfnamefont {C.}~\bibnamefont {Si}}, \ and\ \bibinfo {author} {\bibfnamefont {W.}~\bibnamefont {Duan}},\ }\href {10.1021/acs.nanolett.8b00237} {\bibfield  {journal} {\bibinfo  {journal} {Nano Letters}\ }\textbf {\bibinfo {volume} {18}},\ \bibinfo {pages} {2924} (\bibinfo {year} {2018})}\BibitemShut {NoStop}%
\bibitem [{for()}]{formation_energy}%
  \BibitemOpen
  \href@noop {} {}\bibinfo {note} {The thermodynamic stability of Li adsorption is further confirmed by formation energy ($\Delta$H) calculations relative to bulk Li and pristine NbSe$_2$ monolayer. We placed a Li atom in supercells of varying size-$1 \times 1$, $2 \times 2$, $3 \times 3$, and $4 \times 4$-corresponding to Li concentrations of Li$_1$NbSe$2$, Li$_{0.25}$NbSe$_2$, Li$_{0.1111}$NbSe$_2$, and Li$_{0.0625}$NbSe$_2$, respectively. Our calculations show that all configurations yield negative $\Delta$H, confirming that the Li adsorption is thermodynamically stable. As the Li concentration decreases, the structure becomes relatively more stable with lower in the value of $\Delta$H; however, the energy difference between Li$_{0.1111}$NbSe$_2$ and Li$_{0.0625}$NbSe$_2$ is less than 0.5~meV/atom, indicating negligible energetic variation at low doping levels. Therefore, the $3 \times 3$ supercell is adopted for Li adsorption also to maintain consistency with the periodicity of the distorted CDW
  phase.}\BibitemShut {Stop}%
\bibitem [{\citenamefont {Margine}\ and\ \citenamefont {Giustino}(2013)}]{Margine2013s}%
  \BibitemOpen
  \bibfield  {author} {\bibinfo {author} {\bibfnamefont {E.~R.}\ \bibnamefont {Margine}}\ and\ \bibinfo {author} {\bibfnamefont {F.}~\bibnamefont {Giustino}},\ }\href {\doibase 10.1103/PhysRevB.87.024505} {\bibfield  {journal} {\bibinfo  {journal} {Phys. Rev. B}\ }\textbf {\bibinfo {volume} {87}},\ \bibinfo {pages} {024505} (\bibinfo {year} {2013})}\BibitemShut {NoStop}%
\bibitem [{\citenamefont {Heil}\ \emph {et~al.}(2017)\citenamefont {Heil}, \citenamefont {Ponc\'e}, \citenamefont {Lambert}, \citenamefont {Schlipf}, \citenamefont {Margine},\ and\ \citenamefont {Giustino}}]{Heil2017s}%
  \BibitemOpen
  \bibfield  {author} {\bibinfo {author} {\bibfnamefont {C.}~\bibnamefont {Heil}}, \bibinfo {author} {\bibfnamefont {S.}~\bibnamefont {Ponc\'e}}, \bibinfo {author} {\bibfnamefont {H.}~\bibnamefont {Lambert}}, \bibinfo {author} {\bibfnamefont {M.}~\bibnamefont {Schlipf}}, \bibinfo {author} {\bibfnamefont {E.~R.}\ \bibnamefont {Margine}}, \ and\ \bibinfo {author} {\bibfnamefont {F.}~\bibnamefont {Giustino}},\ }\href {\doibase 10.1103/PhysRevLett.119.087003} {\bibfield  {journal} {\bibinfo  {journal} {Phys. Rev. Lett.}\ }\textbf {\bibinfo {volume} {119}},\ \bibinfo {pages} {087003} (\bibinfo {year} {2017})}\BibitemShut {NoStop}%
\bibitem [{\citenamefont {Giannozzi}\ \emph {et~al.}(2017)\citenamefont {Giannozzi}, \citenamefont {Andreussi}, \citenamefont {Brumme}, \citenamefont {Bunau}, \citenamefont {Nardelli}, \citenamefont {Calandra}, \citenamefont {Car}, \citenamefont {Cavazzoni}, \citenamefont {Ceresoli}, \citenamefont {Cococcioni}, \citenamefont {Colonna}, \citenamefont {Carnimeo}, \citenamefont {Corso}, \citenamefont {de~Gironcoli}, \citenamefont {Delugas}, \citenamefont {DiStasio}, \citenamefont {Ferretti}, \citenamefont {Floris}, \citenamefont {Fratesi}, \citenamefont {Fugallo}, \citenamefont {Gebauer}, \citenamefont {Gerstmann}, \citenamefont {Giustino}, \citenamefont {Gorni}, \citenamefont {Jia}, \citenamefont {Kawamura}, \citenamefont {Ko}, \citenamefont {Kokalj}, \citenamefont {K\"u\c{c}\"ukbenli}, \citenamefont {Lazzeri}, \citenamefont {Marsili}, \citenamefont {Marzari}, \citenamefont {Mauri}, \citenamefont {Nguyen}, \citenamefont {Nguyen}, \citenamefont {de-la Roza}, \citenamefont {Paulatto}, \citenamefont {Ponc\'{e}},
  \citenamefont {Rocca}, \citenamefont {Sabatini}, \citenamefont {Santra}, \citenamefont {Schlipf}, \citenamefont {Seitsonen}, \citenamefont {Smogunov}, \citenamefont {Timrov}, \citenamefont {Thonhauser}, \citenamefont {Umari}, \citenamefont {Vast}, \citenamefont {Wu},\ and\ \citenamefont {Baroni}}]{giannozzi2017o}%
  \BibitemOpen
  \bibfield  {author} {\bibinfo {author} {\bibfnamefont {P.}~\bibnamefont {Giannozzi}}, \bibinfo {author} {\bibfnamefont {O.}~\bibnamefont {Andreussi}}, \bibinfo {author} {\bibfnamefont {T.}~\bibnamefont {Brumme}}, \bibinfo {author} {\bibfnamefont {O.}~\bibnamefont {Bunau}}, \bibinfo {author} {\bibfnamefont {M.~B.}\ \bibnamefont {Nardelli}}, \bibinfo {author} {\bibfnamefont {M.}~\bibnamefont {Calandra}}, \bibinfo {author} {\bibfnamefont {R.}~\bibnamefont {Car}}, \bibinfo {author} {\bibfnamefont {C.}~\bibnamefont {Cavazzoni}}, \bibinfo {author} {\bibfnamefont {D.}~\bibnamefont {Ceresoli}}, \bibinfo {author} {\bibfnamefont {M.}~\bibnamefont {Cococcioni}}, \bibinfo {author} {\bibfnamefont {N.}~\bibnamefont {Colonna}}, \bibinfo {author} {\bibfnamefont {I.}~\bibnamefont {Carnimeo}}, \bibinfo {author} {\bibfnamefont {A.~D.}\ \bibnamefont {Corso}}, \bibinfo {author} {\bibfnamefont {S.}~\bibnamefont {de~Gironcoli}}, \bibinfo {author} {\bibfnamefont {P.}~\bibnamefont {Delugas}}, \bibinfo {author} {\bibfnamefont
  {R.}~\bibnamefont {DiStasio}}, \bibinfo {author} {\bibfnamefont {A.}~\bibnamefont {Ferretti}}, \bibinfo {author} {\bibfnamefont {A.}~\bibnamefont {Floris}}, \bibinfo {author} {\bibfnamefont {G.}~\bibnamefont {Fratesi}}, \bibinfo {author} {\bibfnamefont {G.}~\bibnamefont {Fugallo}}, \bibinfo {author} {\bibfnamefont {R.}~\bibnamefont {Gebauer}}, \bibinfo {author} {\bibfnamefont {U.}~\bibnamefont {Gerstmann}}, \bibinfo {author} {\bibfnamefont {F.}~\bibnamefont {Giustino}}, \bibinfo {author} {\bibfnamefont {T.}~\bibnamefont {Gorni}}, \bibinfo {author} {\bibfnamefont {J.}~\bibnamefont {Jia}}, \bibinfo {author} {\bibfnamefont {M.}~\bibnamefont {Kawamura}}, \bibinfo {author} {\bibfnamefont {H.-Y.}\ \bibnamefont {Ko}}, \bibinfo {author} {\bibfnamefont {A.}~\bibnamefont {Kokalj}}, \bibinfo {author} {\bibfnamefont {E.}~\bibnamefont {K\"u\c{c}\"ukbenli}}, \bibinfo {author} {\bibfnamefont {M.}~\bibnamefont {Lazzeri}}, \bibinfo {author} {\bibfnamefont {M.}~\bibnamefont {Marsili}}, \bibinfo {author} {\bibfnamefont
  {N.}~\bibnamefont {Marzari}}, \bibinfo {author} {\bibfnamefont {F.}~\bibnamefont {Mauri}}, \bibinfo {author} {\bibfnamefont {N.~L.}\ \bibnamefont {Nguyen}}, \bibinfo {author} {\bibfnamefont {H.-V.}\ \bibnamefont {Nguyen}}, \bibinfo {author} {\bibfnamefont {A.~O.}\ \bibnamefont {de-la Roza}}, \bibinfo {author} {\bibfnamefont {L.}~\bibnamefont {Paulatto}}, \bibinfo {author} {\bibfnamefont {S.}~\bibnamefont {Ponc\'{e}}}, \bibinfo {author} {\bibfnamefont {D.}~\bibnamefont {Rocca}}, \bibinfo {author} {\bibfnamefont {R.}~\bibnamefont {Sabatini}}, \bibinfo {author} {\bibfnamefont {B.}~\bibnamefont {Santra}}, \bibinfo {author} {\bibfnamefont {M.}~\bibnamefont {Schlipf}}, \bibinfo {author} {\bibfnamefont {A.~P.}\ \bibnamefont {Seitsonen}}, \bibinfo {author} {\bibfnamefont {A.}~\bibnamefont {Smogunov}}, \bibinfo {author} {\bibfnamefont {I.}~\bibnamefont {Timrov}}, \bibinfo {author} {\bibfnamefont {T.}~\bibnamefont {Thonhauser}}, \bibinfo {author} {\bibfnamefont {P.}~\bibnamefont {Umari}}, \bibinfo {author}
  {\bibfnamefont {N.}~\bibnamefont {Vast}}, \bibinfo {author} {\bibfnamefont {X.}~\bibnamefont {Wu}}, \ and\ \bibinfo {author} {\bibfnamefont {S.}~\bibnamefont {Baroni}},\ }\href {\doibase 10.1088/1361-648X/aa8f79} {\bibfield  {journal} {\bibinfo  {journal} {J. Phys.: Condens. Matter}\ }\textbf {\bibinfo {volume} {29}},\ \bibinfo {pages} {465901} (\bibinfo {year} {2017})}\BibitemShut {NoStop}%
\bibitem [{\citenamefont {van Setten}\ \emph {et~al.}(2018)\citenamefont {van Setten}, \citenamefont {Giantomassi}, \citenamefont {Bousquet}, \citenamefont {Verstraete}, \citenamefont {Hamann}, \citenamefont {Gonze},\ and\ \citenamefont {Rignanese}}]{PseudoDojo2018}%
  \BibitemOpen
  \bibfield  {author} {\bibinfo {author} {\bibfnamefont {M.~J.}\ \bibnamefont {van Setten}}, \bibinfo {author} {\bibfnamefont {M.}~\bibnamefont {Giantomassi}}, \bibinfo {author} {\bibfnamefont {E.}~\bibnamefont {Bousquet}}, \bibinfo {author} {\bibfnamefont {M.~J.}\ \bibnamefont {Verstraete}}, \bibinfo {author} {\bibfnamefont {D.~R.}\ \bibnamefont {Hamann}}, \bibinfo {author} {\bibfnamefont {X.}~\bibnamefont {Gonze}}, \ and\ \bibinfo {author} {\bibfnamefont {G.-M.}\ \bibnamefont {Rignanese}},\ }\href {https://www.sciencedirect.com/science/article/pii/S0010465518300250} {\bibfield  {journal} {\bibinfo  {journal} {Comput. Phys. Commun.}\ }\textbf {\bibinfo {volume} {226}},\ \bibinfo {pages} {39} (\bibinfo {year} {2018})}\BibitemShut {NoStop}%
\bibitem [{\citenamefont {Perdew}\ \emph {et~al.}(1996)\citenamefont {Perdew}, \citenamefont {Burke},\ and\ \citenamefont {Ernzerhof}}]{perdew1996o}%
  \BibitemOpen
  \bibfield  {author} {\bibinfo {author} {\bibfnamefont {J.~P.}\ \bibnamefont {Perdew}}, \bibinfo {author} {\bibfnamefont {K.}~\bibnamefont {Burke}}, \ and\ \bibinfo {author} {\bibfnamefont {M.}~\bibnamefont {Ernzerhof}},\ }\href {\doibase 10.1103/PhysRevLett.77.3865} {\bibfield  {journal} {\bibinfo  {journal} {Phys. Rev. Lett.}\ }\textbf {\bibinfo {volume} {77}},\ \bibinfo {pages} {3865} (\bibinfo {year} {1996})}\BibitemShut {NoStop}%
\bibitem [{\citenamefont {Monkhorst}\ and\ \citenamefont {Pack}(1976)}]{Monkhors1976s}%
  \BibitemOpen
  \bibfield  {author} {\bibinfo {author} {\bibfnamefont {H.~J.}\ \bibnamefont {Monkhorst}}\ and\ \bibinfo {author} {\bibfnamefont {J.~D.}\ \bibnamefont {Pack}},\ }\href {\doibase 10.1103/PhysRevB.13.5188} {\bibfield  {journal} {\bibinfo  {journal} {Phys. Rev. B}\ }\textbf {\bibinfo {volume} {13}},\ \bibinfo {pages} {5188} (\bibinfo {year} {1976})}\BibitemShut {NoStop}%
\bibitem [{\citenamefont {Methfessel}\ and\ \citenamefont {Paxton}(1989)}]{Methfessel1989s}%
  \BibitemOpen
  \bibfield  {author} {\bibinfo {author} {\bibfnamefont {M.}~\bibnamefont {Methfessel}}\ and\ \bibinfo {author} {\bibfnamefont {A.~T.}\ \bibnamefont {Paxton}},\ }\href {\doibase 10.1103/PhysRevB.40.3616} {\bibfield  {journal} {\bibinfo  {journal} {Phys. Rev. B}\ }\textbf {\bibinfo {volume} {40}},\ \bibinfo {pages} {3616} (\bibinfo {year} {1989})}\BibitemShut {NoStop}%
\bibitem [{\citenamefont {Baroni}\ \emph {et~al.}(2001)\citenamefont {Baroni}, \citenamefont {de~Gironcoli}, \citenamefont {Corso},\ and\ \citenamefont {Giannozzi}}]{Baroni2001s}%
  \BibitemOpen
  \bibfield  {author} {\bibinfo {author} {\bibfnamefont {S.}~\bibnamefont {Baroni}}, \bibinfo {author} {\bibfnamefont {S.}~\bibnamefont {de~Gironcoli}}, \bibinfo {author} {\bibfnamefont {A.~D.}\ \bibnamefont {Corso}}, \ and\ \bibinfo {author} {\bibfnamefont {P.}~\bibnamefont {Giannozzi}},\ }\href {\doibase 10.1103/RevModPhys.73.515} {\bibfield  {journal} {\bibinfo  {journal} {Rev. Mod. Phys.}\ }\textbf {\bibinfo {volume} {73}},\ \bibinfo {pages} {515} (\bibinfo {year} {2001})}\BibitemShut {NoStop}%
\bibitem [{\citenamefont {Allen}\ and\ \citenamefont {Dynes}(1975)}]{Allen1975s}%
  \BibitemOpen
  \bibfield  {author} {\bibinfo {author} {\bibfnamefont {P.~B.}\ \bibnamefont {Allen}}\ and\ \bibinfo {author} {\bibfnamefont {R.~C.}\ \bibnamefont {Dynes}},\ }\href {\doibase 10.1103/PhysRevB.12.905} {\bibfield  {journal} {\bibinfo  {journal} {Phys. Rev. B}\ }\textbf {\bibinfo {volume} {12}},\ \bibinfo {pages} {905} (\bibinfo {year} {1975})}\BibitemShut {NoStop}%
\bibitem [{\citenamefont {Lee}\ \emph {et~al.}(2023)\citenamefont {Lee}, \citenamefont {Ponc{\'e}}, \citenamefont {Bushick}, \citenamefont {Hajinazar}, \citenamefont {Lafuente-Bartolome}, \citenamefont {Leveillee}, \citenamefont {Lian}, \citenamefont {Lihm}, \citenamefont {Macheda}, \citenamefont {Mori}, \citenamefont {Paudyal}, \citenamefont {Sio}, \citenamefont {Tiwari}, \citenamefont {Zacharias}, \citenamefont {Zhang}, \citenamefont {Bonini}, \citenamefont {Kioupakis}, \citenamefont {Margine},\ and\ \citenamefont {Giustino}}]{EPW2023}%
  \BibitemOpen
  \bibfield  {author} {\bibinfo {author} {\bibfnamefont {H.}~\bibnamefont {Lee}}, \bibinfo {author} {\bibfnamefont {S.}~\bibnamefont {Ponc{\'e}}}, \bibinfo {author} {\bibfnamefont {K.}~\bibnamefont {Bushick}}, \bibinfo {author} {\bibfnamefont {S.}~\bibnamefont {Hajinazar}}, \bibinfo {author} {\bibfnamefont {J.}~\bibnamefont {Lafuente-Bartolome}}, \bibinfo {author} {\bibfnamefont {J.}~\bibnamefont {Leveillee}}, \bibinfo {author} {\bibfnamefont {C.}~\bibnamefont {Lian}}, \bibinfo {author} {\bibfnamefont {J.-M.}\ \bibnamefont {Lihm}}, \bibinfo {author} {\bibfnamefont {F.}~\bibnamefont {Macheda}}, \bibinfo {author} {\bibfnamefont {H.}~\bibnamefont {Mori}}, \bibinfo {author} {\bibfnamefont {H.}~\bibnamefont {Paudyal}}, \bibinfo {author} {\bibfnamefont {W.~H.}\ \bibnamefont {Sio}}, \bibinfo {author} {\bibfnamefont {S.}~\bibnamefont {Tiwari}}, \bibinfo {author} {\bibfnamefont {M.}~\bibnamefont {Zacharias}}, \bibinfo {author} {\bibfnamefont {X.}~\bibnamefont {Zhang}}, \bibinfo {author} {\bibfnamefont
  {N.}~\bibnamefont {Bonini}}, \bibinfo {author} {\bibfnamefont {E.}~\bibnamefont {Kioupakis}}, \bibinfo {author} {\bibfnamefont {E.~R.}\ \bibnamefont {Margine}}, \ and\ \bibinfo {author} {\bibfnamefont {F.}~\bibnamefont {Giustino}},\ }\href {https://doi.org/10.1038/s41524-023-01107-3} {\bibfield  {journal} {\bibinfo  {journal} {npj Comput. Mater.}\ }\textbf {\bibinfo {volume} {9}},\ \bibinfo {pages} {156} (\bibinfo {year} {2023})}\BibitemShut {NoStop}%
\bibitem [{\citenamefont {Pizzi}\ \emph {et~al.}(2019)\citenamefont {Pizzi}, \citenamefont {Vitale}, \citenamefont {Arita}, \citenamefont {Blügel}, \citenamefont {Freimuth}, \citenamefont {G{\'{e}}ranton}, \citenamefont {Gibertini}, \citenamefont {Gresch}, \citenamefont {Johnson}, \citenamefont {Koretsune}, \citenamefont {Iba{\~{n}}ez-Azpiroz}, \citenamefont {Lee}, \citenamefont {Lihm}, \citenamefont {Marchand}, \citenamefont {Marrazzo}, \citenamefont {Mokrousov}, \citenamefont {Mustafa}, \citenamefont {Nohara}, \citenamefont {Nomura}, \citenamefont {Paulatto}, \citenamefont {Ponc{\'{e}}}, \citenamefont {Ponweiser}, \citenamefont {Qiao}, \citenamefont {Thöle}, \citenamefont {Tsirkin}, \citenamefont {Wierzbowska}, \citenamefont {Marzari}, \citenamefont {Vanderbilt}, \citenamefont {Souza}, \citenamefont {Mostofi},\ and\ \citenamefont {Yates}}]{Pizzi2019s}%
  \BibitemOpen
  \bibfield  {author} {\bibinfo {author} {\bibfnamefont {G.}~\bibnamefont {Pizzi}}, \bibinfo {author} {\bibfnamefont {V.}~\bibnamefont {Vitale}}, \bibinfo {author} {\bibfnamefont {R.}~\bibnamefont {Arita}}, \bibinfo {author} {\bibfnamefont {S.}~\bibnamefont {Blügel}}, \bibinfo {author} {\bibfnamefont {F.}~\bibnamefont {Freimuth}}, \bibinfo {author} {\bibfnamefont {G.}~\bibnamefont {G{\'{e}}ranton}}, \bibinfo {author} {\bibfnamefont {M.}~\bibnamefont {Gibertini}}, \bibinfo {author} {\bibfnamefont {D.}~\bibnamefont {Gresch}}, \bibinfo {author} {\bibfnamefont {C.}~\bibnamefont {Johnson}}, \bibinfo {author} {\bibfnamefont {T.}~\bibnamefont {Koretsune}}, \bibinfo {author} {\bibfnamefont {J.}~\bibnamefont {Iba{\~{n}}ez-Azpiroz}}, \bibinfo {author} {\bibfnamefont {H.}~\bibnamefont {Lee}}, \bibinfo {author} {\bibfnamefont {J.-M.}\ \bibnamefont {Lihm}}, \bibinfo {author} {\bibfnamefont {D.}~\bibnamefont {Marchand}}, \bibinfo {author} {\bibfnamefont {A.}~\bibnamefont {Marrazzo}}, \bibinfo {author} {\bibfnamefont
  {Y.}~\bibnamefont {Mokrousov}}, \bibinfo {author} {\bibfnamefont {J.~I.}\ \bibnamefont {Mustafa}}, \bibinfo {author} {\bibfnamefont {Y.}~\bibnamefont {Nohara}}, \bibinfo {author} {\bibfnamefont {Y.}~\bibnamefont {Nomura}}, \bibinfo {author} {\bibfnamefont {L.}~\bibnamefont {Paulatto}}, \bibinfo {author} {\bibfnamefont {S.}~\bibnamefont {Ponc{\'{e}}}}, \bibinfo {author} {\bibfnamefont {T.}~\bibnamefont {Ponweiser}}, \bibinfo {author} {\bibfnamefont {J.}~\bibnamefont {Qiao}}, \bibinfo {author} {\bibfnamefont {F.}~\bibnamefont {Thöle}}, \bibinfo {author} {\bibfnamefont {S.~S.}\ \bibnamefont {Tsirkin}}, \bibinfo {author} {\bibfnamefont {M.}~\bibnamefont {Wierzbowska}}, \bibinfo {author} {\bibfnamefont {N.}~\bibnamefont {Marzari}}, \bibinfo {author} {\bibfnamefont {D.}~\bibnamefont {Vanderbilt}}, \bibinfo {author} {\bibfnamefont {I.}~\bibnamefont {Souza}}, \bibinfo {author} {\bibfnamefont {A.~A.}\ \bibnamefont {Mostofi}}, \ and\ \bibinfo {author} {\bibfnamefont {J.~R.}\ \bibnamefont {Yates}},\ }\href {\doibase
  10.1088/1361-648X/ab51ff} {\bibfield  {journal} {\bibinfo  {journal} {J. Phys.: Condens. Matter.}\ }\textbf {\bibinfo {volume} {32}},\ \bibinfo {pages} {165902} (\bibinfo {year} {2019})}\BibitemShut {NoStop}%
\end{thebibliography}%
\end{document}